\documentclass[array]{aastex701}
\usepackage{graphicx}        
\usepackage{color}           
\usepackage{amsmath}         
\usepackage{rotating} 
\usepackage{tabularx}
\usepackage{hyperref}
\usepackage{booktabs}
\usepackage{longtable}
\newcommand{\ie}{{i.e.}}     

\chardef\us=`\_
\begin{document}
\title{Resolving the Tachocline using Inversion of Rotational Splitting 
Derived from Fitting Very Long and Long Time Series}
\author[orcid=0000-0003-1531-1541, sname=Korzennik, gname=Sylvain]{Sylvain G. Korzennik}
\affiliation{Center for Astrophysics $|$ Harvard \& Smithsonian, Cambridge, MA 02138, USA}
\email[show]{skorzennik@cfa.harvard.edu}
\correspondingauthor{Sylvain G. Korzennik}

\author{Antonio Eff-Darwich}
\affiliation{Universidad de La Laguna, Tenerife, 38204, Spain}
\affiliation{Instituto de Astrof\'isica de Canarias, Tenerife, 38206, Spain} 
\email[show]{adarwich@ull.edu.es}

\begin{abstract}

We use rotation splittings derived from very long and long time series, namely
25.2, 12.6 and 6.3 year long, computed by \citet{Korzennik-2023} independent
methodology to characterize the solar tachocline and its variation with
latitude and time.
We use two different inversion methodologies and a model of the tachocline to
derive its position, width and the amplitude of the radial shear.
To validate our methodology we present results from simulated rotational
splittings, whether including or not random noise commensurable with the
current observational precision. We also describe how we leverage the fact
that one of our methodologies uses an initial guess that can be chosen to
include {\em a priori} information.
In order to try to resolve the tachocline, we increased the radial density of
the inversion grid and showed how it affect the inferences.
We also show how the trade off between smoothing and noise magnification
affects these, as well as the effectiveness of using an informed initial
guess.
Results derived from high-precision rotational splittings show clearly that
the location of the tachocline at low latitudes is different for its position
at high latitudes.
The latitudinal variation of its width is not significantly constrained, but
our results agree with estimates based on forward modeling.
When using splittings derived from somewhat shorter time series, we find 
temporal variations that are neither definitive nor significant, since we
see systematic differences when using different methodologies.
  
\end{abstract}


%
%
\section{Introduction}\label{S-Introduction} 

The solar tachocline, a term introduced by \citet{spiegel1992}, is the shear
layer where the solar rotation transitions from a nearly rigid-body rotation
in the radiative interior to a differential rotation in the convection zone,
\ie, where the rotation varies with depth and latitude. This transition layer
is located near the base of the convective envelope, \ie, approximately at
$0.7\,R_\odot$, and is believed to be the seat of the solar dynamo, the
mechanism responsible for generating the Sun's magnetic field. It is also
believed to be responsible for the solar activity cycles and to play a vital
role in angular momentum transport and chemical mixing within the Sun
\citep{strugaret2023}.

Over the past three decades, helioseismic analyses of the internal solar
rotation rate \citep{thompson2003}, based on p-mode oscillation frequencies
and employing both global and local inversion techniques, have provided
increasingly precise estimates of the tachocline's location, thickness, and
temporal variability (see Table~\ref{tab:rot-tacho-full}). These studies rely
on observations from instruments such as BBSO \citep{libbrecht1989}, SDO-HMI
\citep{schou2012}, SOI-MDI \citep{scherrer1995}, GONG \citep{harvey1996}, and
LOWL \citep{tomczyk1995}, which have provided frequency measurements of
acoustic modes whose internal turning points are distributed both above and
below the tachocline, thereby enabling sensitive diagnostics of this
rotational shear layer. Helioseismic results place the tachocline between
$0.69$ and $0.72\,R_\odot$, with estimates of its thickness at the equator
ranging from $0.006$ to $0.05\,R_\odot$ (see
Table~\ref{tab:rot-tacho-sum}). Despite general agreement on its location,
substantial variation persists across studies, particularly regarding the
tachocline's latitudinal profile and temporal evolution.
Some studies suggest a prolate geometry, \ie, thicker and deeper at higher
latitudes, while others suggest a more spherical or indeterminate
structure. Similarly, while several analyses detect temporal changes in the
amplitude of the shear with respect to time, hence the solar activity level,
others find no significant variation.

These discrepancies arise not only from the limited spatial resolution of
helioseismic inversions, particularly near the tachocline, but also from
differences in the length and quality of the input data sets, the inversion
methodologies, and the parameterizations used to define the tachocline
properties.
In any case, it is clear that the tachocline is a dynamic, spatially complex
region whose precise characterization remains challenging. Hence, it is
necessary to implement inversion techniques capable of resolving sharp
gradients and fine-scale features while minimizing smoothing, typically used
to alleviate the intrinsic singular nature of the rotation inverse problem,
while limiting noise magnification. Moreover, applying such techniques
consistently across diverse data sets derived from different instruments is
essential for building a coherent and physically meaningful picture of the
tachocline.

\begin{sidewaystable}
\centering
\small
\begin{tabular}{l c c l l}
\toprule
\textbf{Instrument}  & \textbf{Location} & \textbf{Width} & \textbf{Observed Variation} & \textbf{Reference}  \\ \midrule
 BBSO           & 0.692 $\pm$ 0.005            & 0.096 $\pm$ 0.04                     &                             & Kosovichev (1996)   \\ \midrule
 Mean GONG+BBSO & 0.7050 $\pm$ 0.0027          & 0.0098 $\pm$ 0.0026                  &                             & Basu (1997)         \\ \midrule
 GONG           &  0.675 -- 0.695              & 0.001 -- 0.013                       & Hints of latitude dependence, \\
                &                              &                                      & not statistically significant. Dependence on the analysis technique
                                                                                                                    & Antia, Basu \& Chitre (1998) \\ \midrule
 LOWL & 0.695 $\pm$ 0.005                      &  0.05 $\pm$ 0.03                     & (half-width) ~              & Corbard et al. (1998)        \\ \midrule 
 LOWL & 0.701 $\pm$ 0.004                      & $<$0.05                              & ~                           & Corbard et al. (1999)        \\ \midrule
 LOWL & 0.693 $\pm$ 0.002                      & 0.039 $\pm$ 0.013                    & Prolate profile: radius increases by $\sim$0.024 $\pm$ 0.004 $R_\odot$ \\
      &                                        &                                      & from equator to $60^\circ$. No significant variation of thickness with latitude
                                                                                                                    & Charbonneau et al. (1999)    \\ \midrule
 MDI  & 0.697 $\pm$ 0.002                      & 0.019 $\pm$ 0.001                    & Based on inversions of the sound speed
                                                                                                                    & Elliott \& Gough (1999)      \\ \midrule
      & Not specified                          & Not specified                        & Rotation rate increases with depth at low/mid latitudes, \\
      &                                        &                                      & 1.3-year oscillations near equator, $\sim1$ year at high latitudes
                                                                                                                    & Howe et al.~(2000)  \\ \midrule
 Mean GONG+MDI & 0.6893 $\pm$ 0.0023           & 0.0061 $\pm$ 0.0012                  & Clear latitudinal variation in depth. \\ & & &  No significant temporal variations
                                                                                                                    & Basu \& Antia (2001) \\  \midrule
 Mean GONG+MDI & 0.6916 $\pm$ 0.0019           & 0.0065 $\pm$ 0.0013                  & Clear latitudinal variation in depth and width. \\ & & & No significant temporal variations
                                                                                                                    & Basu \& Antia (2003) \\ \midrule
 GONG & 0.6945 $\pm$ 0.0013                    & 0.0037 $\pm$ 0.0011                  & Center is prolate and thickness increases with latitude. \\ 
 MDI  & 0.6915 $\pm$ 0.0016                    & 0.0028 $\pm$ 0.0017                  & Clear temporal variation of $\delta\Omega$ with solar activity
                                                                                                                    & Antia \& Basu (2011) \\ \midrule 
 GONG & 0.7067 –- 0.7187                       & 0.0067–0.0198                        \\
 MDI  & 0.708  –- 0.7181                       & 0.003  –- 0.0188                     & Clear temporal variation across solar cycles 23–25, \\
 HMI  & 0.7058 –- 0.7166                       & 0.0077 –- 0.0182                     & changes in depth and $\delta\Omega$ 
                                                                                                                    & Basu et al. (2024)   \\ \bottomrule
\end{tabular}
\caption{Summary of tachocline properties from various helioseismic
  studies. Observed variations include only latitudinal or temporal
  effects when reported.}
\label{tab:rot-tacho-full}
\end{sidewaystable}

\begin{table}[htbp]
\begin{tabular}{lcc} \toprule
\textbf{Instrument} & \textbf{Position} & \textbf{Width}   \\ \midrule
MDI                 & 0.692 -- 0.718    & 0.003 -– 0.072   \\
GONG                & 0.689 -- 0.719    & 0.006 -- 0.096   \\
LOWL                & 0.690 -- 0.700    & 0.026 -- 0.080   \\
BBSO                & 0.690 -- 0.696    & 0.031 -- 0.080   \\
HMI                 & 0.706 -- 0.717    & 0.008 -- 0.018 \\ \bottomrule
\end{tabular}
\caption{Summary of the position and width of the tachocline at the
  equator as estimated from each instrument.}
\label{tab:rot-tacho-sum}
\end{table}

Although the present work focuses primarily on the internal rotation rate of
the Sun, complementary and essential insights are provided by helioseismic
inferences of the Sun's internal structure, particularly those deriving the
difference in sound speed between observations and the standard solar model in
the vicinity of the tachocline. Structural inversions of helioseismic data
reveal a distinct bump in the sound speed profile near $0.65\,R_\odot$,
coinciding with the location of the tachocline, where observed values exceed
predictions from the standard solar model. This anomaly is generally
attributed to compositional differences in the tachocline, particularly the
helium mixing that reduces the local mean molecular weight and thereby
increases the sound speed \citep{gough1998}. These structural signatures
provide independent constraints on the depth, composition, and mixing
processes within the tachocline, and therefore offer valuable complementary
results to those derived from rotation inversions \citep{eliot1999,
  takata2003}.

A number of theoretical models have been developed to explain the formation,
stability, and confinement of the tachocline --- \ie, its remarkable thinness
--- each emphasizing different physical mechanisms (see a sample of them in
Table~\ref{tab:model-1}). Hydrodynamic models (e.g., \citealt{spiegel1992})
propose viscous shear spreading, constrained by anisotropic turbulence, as
such mechanism, whereas magnetic confinement models rely on either fossil
fields resisting differential shear (\citealt{gough1998}) or dynamo-generated
oscillatory fields (\citealt{forgacs2001}). Additional models highlight the
role of thermal wind balance, meridional circulation, and baroclinic forcing
in shaping the tachocline’s structure (\citealt{rempel2005,
  balbus2010}). Predictions for the tachocline's location typically place it
near the base of the convection zone (around $0.69\,R_\odot$), with
model-dependent thicknesses ranging from very thin ($\sim0.02\,R_\odot$) to
more extended ($\sim0.1\,R_\odot$). Several models account for latitudinal
variation, often predicting a prolate shape (thicker at higher latitudes), and
some suggest temporal modulations tied to the solar magnetic cycle or internal
dynamo dynamics. While steady-state structures are assumed in some cases,
others predict cyclic or irregular variability, reflecting the dynamic
interplay between rotational shear, magnetic fields, and turbulent mixing.

Hence, it is essential to improve the observational constraints provided by
helioseismic inversions, including isorotation contours, detailed latitudinal
and radial shear gradients, confirming the nearly uniform rotation of the
radiative interior, and measuring the tachocline's position and thickness as
well as their latitudinal variation. These characteristics are essential for
understanding its long-term stability against radiative spreading. They are
also vital for constraining the underlying angular momentum transport
mechanism, like the nature of turbulent anisotropy (e.g., horizontal
vs.\ radial efficiency) and the effectiveness of magnetic confinement by
either fossil or dynamo-generated oscillatory fields. Finally, long-term,
continuous helioseismic observations with high spatial resolution are required
to detect subtle temporal variations of the tachocline potentially linked to
the solar cycle or internal dynamo dynamics.

\begin{sidewaystable}[htbp]
\centering
\caption{Summary of some theoretical tachocline confinement models
  and their predictions.}
\label{tab:tachocline-summary}
\begin{tabular}{l l l}
\toprule
\textbf{Confinement Mechanism}                                                & \textbf{Properties, if given}                                 & \textbf{Reference}  \\ \midrule
Confinement due to highly anisotropic turbulent viscosity                     & Width $\sim 0.13$ to $0.1\,R_\odot$                             & \cite{spiegel1992} \\
  with a large horizontal component                                           &                                                                & \cite{eliot1997}   \\ \hline
Magnetic field quenches shear in RZ                                           & Thin implied; thickness increases towards the poles            & \cite{rudiger1997} \\ \hline
Fossil magnetic field suppresses shear in RZ                                  & Thin                                                           & \cite{gough1998}   \\ 
  explains rigid interior rotation                                            & (i.e., below helioseismic inversion spatial resolution)        & \cite{garaud2008}  \\ \hline
Turbulent tachocline pervaded and                                             & Thin                                                           & \\
   confined by a dynamo-generated field                                       &                                                                & \cite{forgacs2001} \\ \hline
Anisotropy in momentum transport                                              & Implied thinness \\
   due to large horizontal turbulent viscosity                                &                                                                & \cite{leprovost2006} \\ \hline
Global dynamical model,                                                       & Base convection zone \\
   but is not able to avoid tachocline spreading with time                    &                                                                & \cite{brun2011} \\ \hline
Baroclinic radiative equilibrium                                              & Thin and sharp layer \\
   with no significant meridional circulation for stable confined tachocline  &                                                                      & \cite{caleo2015} \\ \hline
This model does not explain tachocline confinement                            & Much thicker above $50^\circ$ \\ & to avoid hydrodynamic instability  & \cite{gilman2018} \\ \hline
Persistent nonaxisymmetric dynamo field                                       & Width $<0.05\,R_\odot$ \\
    regulates tachocline shear via Maxwell stresses                           &                                                                       & \cite{matilsky2024} \\
\bottomrule
\end{tabular}
\label{tab:model-1}
\end{sidewaystable}

In summary, we need to move beyond broad qualitative descriptions and toward a
quantitatively constrained characterization of the tachocline’s structure and
its role in solar dynamics. This requires not only enhanced spatial
resolution, sensitivity, and temporal coverage, but also the systematic
application of consistent and robust inversion techniques across multiple data
sets and instruments.

\section{Data Sets}\label{S-Data}

In order to best characterize the solar tachocline using rotation inversions
we want to use rotational splittings with the lowest possible uncertainties,
hence values estimated using the longest available time series.
In contrast to the instruments' respective reduction pipe-lines, observations
taken by MDI, HMI and GONG have been fitted by \citet{Korzennik-2023} using
time series of various lengths. Namely, sets as short as 36 days, but also
sets of 72, 144, 288, etc, up to 2,304 days and 4,608 days (\ie, $32\times$
and $64\times72$ days, or 6.3 and 12.6 years respectively) for all 3
instruments, and up to 9,216 days (\ie, $128\times72$ days, or 25.2 years) for
GONG.
For time series longer than 72 days, the fitted epochs overlap by 50\%, while
their starting time is offset by an integer multiple of 72 days with respect
to 1996.05.01 (\ie, the fist day MDI produced science quality data).

\begin{figure}
  \begin{center}
    \includegraphics[width=0.85\textwidth]{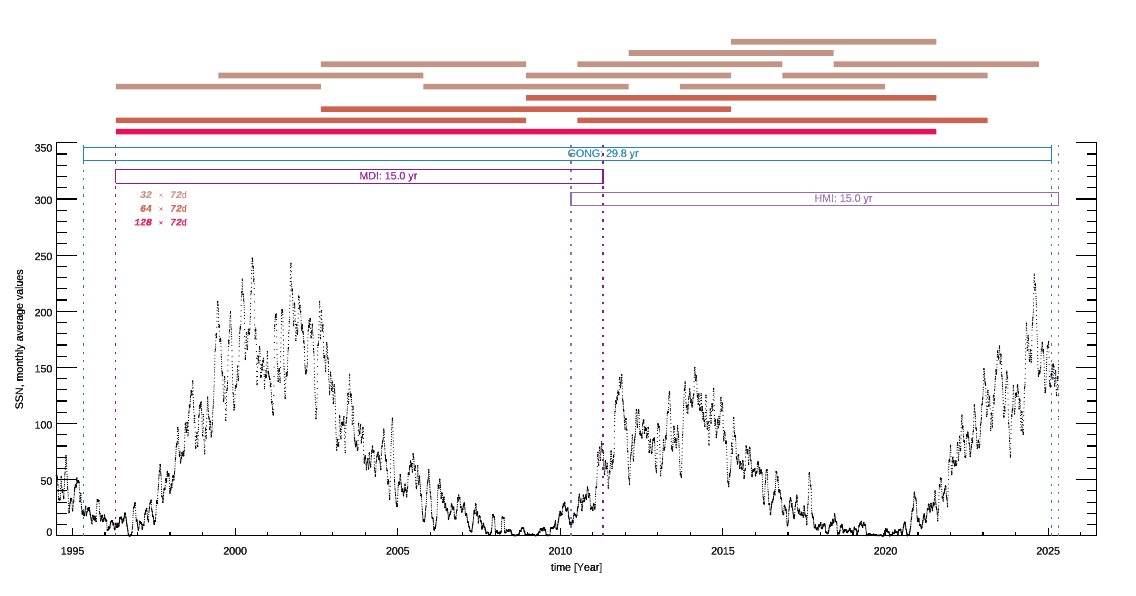} 
  \end{center}
  \caption{Temporal coverage of the time series used in the work presented
    here (filled boxes), compared the temporal coverage of the observations
    (open boxes) for the MDI, HMI and GONG instruments, and to the solar
    activity illustrated by the monthly averaged sunspot number.}
  \label{fig:temp-coverage}
\end{figure}

To attempt resolving the tachocline, we opted to use rotational splittings
resulting from fitting a very long (\ie, $128\times72$ day) and long (\ie,
$32\times$ and $64\times72$ day) time series.
The temporal coverage of these time series, compared with that of the three
instruments, as well as the solar activity, represented by the sunspot number,
are plotted in Fig.~\ref{fig:temp-coverage}.

\section{Methodology and Inverse Techniques}\label{S-Methods}

\subsection{Variable Radial Gridding}

Since we try to resolve the tachocline, we decided to break from tradition and
use a model grid -- \ie, the set of target locations -- with a high resolution
in the radial direction, especially around the tachocline. The resolution in
latitude was in all cases left uniform.
Since our inversion techniques (see below) allow for a variable grid spacing,
we experiment with various grids. Of course, the simplest grid is a uniformly
spaced grid, but in order to reach a high resolution at the tachocline (i.e.,
well below 0.5\% in the radial direction) the grid size becomes large, hence
the size of the problem (\ie, the sizes of the matrices to handle) gets very
large, requiring large amounts of memory and quite long computation time. With
these constraints in mind, we defined such a grid using 256 radial points and
54 latitudinal points, or 13,824 target locations. The radial resolution of
this grid is 0.39\%.

Alternatively, one can change the grid spacing to be large in the radiative
zone (\ie, below the tachocline), where one knows from past experience that
the radial resolution of rotation inversions is low while the solar rotation
rate does not appear to vary much with respect to radius. One can also set the
spacing above the tachocline (\ie, in the convection zone) to a value small
enough to resolve the known radial variation on that region, while using, in a
radial range centered around the tachocline, a much smaller spacing. To avoid
edge effects, we realized through trial and error, that it is best if the
second derivative of the radial grid remains smooth. Following these
guidelines, we defined what we call an {\em aggressive} grid, a grid that
reaches at the tachocline a resolution of 0.09\% while remaining a set of 97
by 56, or only 5,432 target locations.

Finally, we also defined what we call {\em progressive} grids, where the
radial spacing was set to vary more smoothly, \ie, using a polynomial in $r$,
with two target resolutions at the tachocline of 0.25\% and 0.18\%, hence a
set of 142 by 56 or 7,952 target points and a set of 171 by 56 or 9,576 target
points.

\begin{figure}
  \begin{center}
    \includegraphics[width=0.85\textwidth]{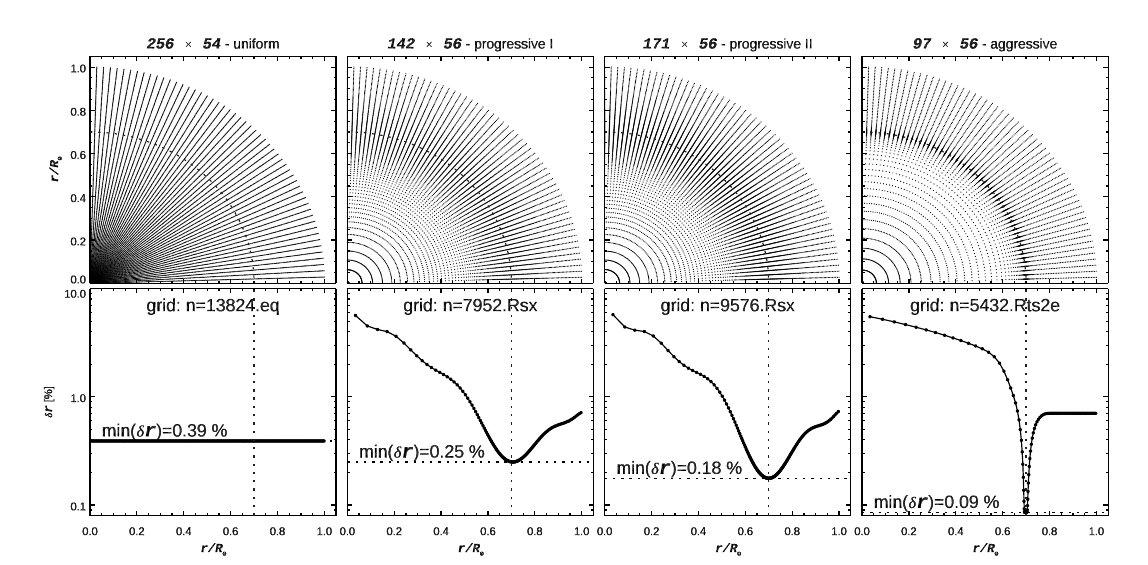} 
  \end{center}
  \caption{Positions of the target locations, shown in Cartesian coordinates
    (top panels) and the grids' radial spacing as a function of radius (bottom
    panels) for the four grids we used. Shown, from left to right, are the
    uniform radial spacing grid, two {\em progressive} radial spacing grids
    and the so-called {\em aggressive} radial spacing grid.  The lowest radial
    resolution, occurring around the tachocline, is indicated as well.}
  \label{fig:grids}   
\end{figure}

These 4 grids are illustrated in Fig.~\ref{fig:grids} where the positions of
the target locations are shown in Cartesian coordinates, as well as the
corresponding grid radial spacing as a function of radius.

\subsection{Inversion Techniques}

We use two different inversion methods: a variant of the classical regularized
least squares technique \cite[RLS, see][]{EffDarwichEtal-1997} and a newly
developed iterative technique based on the simultaneous algebraic
reconstruction technique \cite[SART,][]{Korzennik+EffDarwich-2024}.  The SART
technique was originally implemented to solve linear systems in image
reconstruction \citep{Anderson+Kak-1984,Sluis-1990} and was proposed as an
inversion technique for the solar internal rotation rate by
\citet{EffDarwich-2010}.

Both methods use a piecewise constant discretization of the two-dimensional
rotation inverse problem, and their implementation allows for a non-uniform
discretization grid (aka target locations). Both are coded to allow the
computation of the averaging kernels, $K^{(a)}_{i,j}(r, \theta)$.
Indeed, the two-dimensional rotation inverse problem can be written as
\begin{equation}
  \frac{\delta\nu_{n,\ell,m}}{2\,m} = \iint K^{(r)}_{n,\ell,m}(r, \theta)\, \Omega(r, \theta)\, dr\, d\theta
\end{equation}
where $\delta \nu_{n,\ell,m} = \nu_{n,\ell,m} - \nu_{n,\ell,-m} $ are the
individual rotation splittings, $\nu_{n,\ell,m}$ the frequency for the mode
with radial order $n$, spherical harmonic degree $\ell$, and azimuthal order
$m$. $K^{(r)}_{n,\ell,m}$ is the corresponding rotation kernel
\cite[see][]{Hansen-1977,EffDarwichEtal-1997,Korzennik+EffDarwich-2024},
$\Omega$ the rotation rate, $r$ the radius, and $\theta$ the co-latitude.

Notice that we use the individual rotation splittings, not expansion
coefficients of these splittings as a function of $m$, since (i) the tables
resulting from the fitting by \citet{Korzennik-2023} provide these individual
frequencies from which we compute the splittings; and (ii) using individual
splittings results in better angular resolution since the observables are not
pre-smoothed to an arbitrary number of expansion coefficients.

When that inverse problem is solved as a linear combination of the splittings,
the resulting estimate of the rotation rate, $\tilde{\Omega}$, is a
convolution of the {\em true} underlying rotation rate, $\Omega$, by a linear
combination of the rotational kernels. That linear combination is the
averaging kernel, namely:

\begin{eqnarray}
  \tilde{\Omega}(r_i, \theta_j) & = & \sum_{n,\ell,m} c^{(i,j)}_{n,\ell,m} \frac{\delta\nu_{n,\ell,m}}{2\,m} \\
     & = & \iint \sum_{n,\ell.m} c^{(i,j)}_{n,\ell,m} K^{(r)}_{n,\ell,m}(r, \theta)\, \Omega(r, \theta)\, dr\, d\theta \\
     & = & \iint K^{(a)}_{(i, j)}(r, \theta)\, \Omega(r, \theta)\, dr\, d\theta
\end{eqnarray}
hence
\begin{equation}
    K^{(a)}_{(i, j)}(r, \theta) = \sum_{n,\ell.m} c^{(i,j)}_{n,\ell,m} K^{(r)}_{n,\ell,m}(r, \theta)
\end{equation}

Our RLS technique is computationally fast, and we implemented it with a single
trade-off parameter (aka Lagrangian multiplier) that controls the amount of
smoothing on the solution by minimizing its second derivative. In contrast,
the SART technique is slow, since it is an iterative method, and we
implemented it using either a global or local smoothing. The local smoothing
is a parametrized function of depth and latitude \citep[see][for additional
  details]{Korzennik+EffDarwich-2024}. Since there is some commonality in the
pre-processing of both methods, i.e., the discretization of matrices on the
model grid, that common first step is coded as a separate computation, and is
parallelized using the message passing interface (MPI) to speed it up. The
second step of the SART inversion is also parallelized using MPI, while the
second step of the RLS is not since it is relatively fast.

Finally, since the SART technique starts with an initial guess, it gives us
the option to use either a simple starting value, like a constant close to the
mean rotation rate, or an informed initial guess, like a two-dimensional
approximation of the anticipated solution.

\subsection{SART Initial Guess}

The simplest way to define the SART initial value is to start with a constant
value at each grid point, picking a value close to the mean rotation. Hence we
selected a starting value of 470 nHz. We also checked that we run enough
iterations for the technique to converge, by saving the solution at
intermediate number of iterations and confirming that the changes of the
solution become negligible.

We also investigated whether we can exploit the need for a starting value by
using an informed guess. If we assume that the solution obtained with the RLS
technique is a smoothed version of the true rotation, we should be able to use
it to produce such informed initial guess. Since we are interested in the
properties of the tachocline, and specifically its width, let us assume that
the RLS solution is a good initial guess, but should be sharpened, since the
RLS regularization is smoothing of the true solution, while the inferred
position of the tachocline does not need any adjustment.

Our current implementation of such sharpening is to fit a sigmoid to the
tachocline, assume that its location is correct, but that its width is
overestimated. Therefore we replace a portion of the RLS solution by a more
narrow sigmoid, typically reducing its width by a factor two, and using this
narrow profile as an initial guess for the SART technique.

As presented in the next section, we used simulated splittings computed from
model rotation profiles to not only test how well our techniques recover the
tachocline, but also to validate our SART initial guess approach.

\section{Results from Simulations}\label{S-R-Simulations}

\begin{figure}
  \begin{center}
    \includegraphics[width=0.9\textwidth]{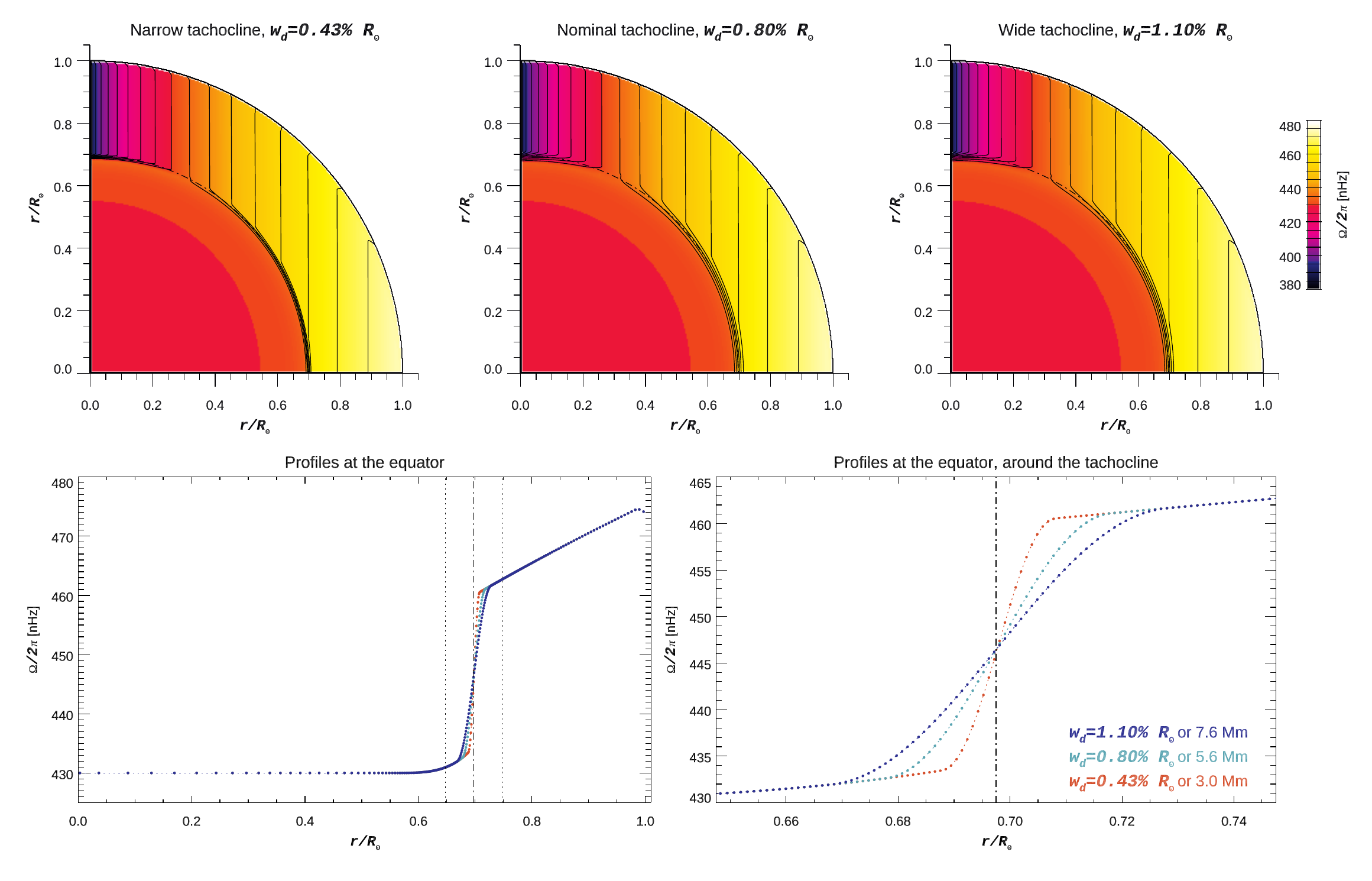} 
  \end{center}
  \caption{Three models used for our simulation. The top panels show the
    simulated rotation rate in Cartesian coordinates, while the bottom panels
    show the rotation profile at the equator, with a zoomed view around the
    tachocline in the lower left panel.} %
  \label{fig:models}
\end{figure}

We present results from simulated splittings computed from simple but
realistic rotational models. Namely a solid body rotation below the
tachocline, a differential rotation above the tachocline, although constant on
cylinder (to use a simple parametrization) and a tachocline whose width is
different for different models. From these models, we compute the individual
rotational splittings and assign to each an uncertainty corresponding to the
error bar for that mode derived from fitting a $32 \times 72$ day long time
series. Assigning a realistic relative uncertainty is key to replicating
adequately the resolution potential of observations, even when inverting
noiseless splittings.
These three models are illustrated in Fig.~\ref{fig:models}, for a so-called
{\em wide}, {\em nominal}, and {\em narrow} tachocline.

Of course, to estimate the effect of observational noise, we also added random
noise to each splitting. This noise is scaled by the mode's uncertainty and an
arbitrary factor. We used the values 0.1, 0.5 and 1.0 for that extra factor,
simulating random errors corresponding to fitting a $100\times, 4\times,$ or
$1\times$ ($32\times 72$-day) long time series.

\subsection{Noiseless Simulations}

\begin{figure}
  \begin{center}
    \includegraphics[width=0.9\textwidth]{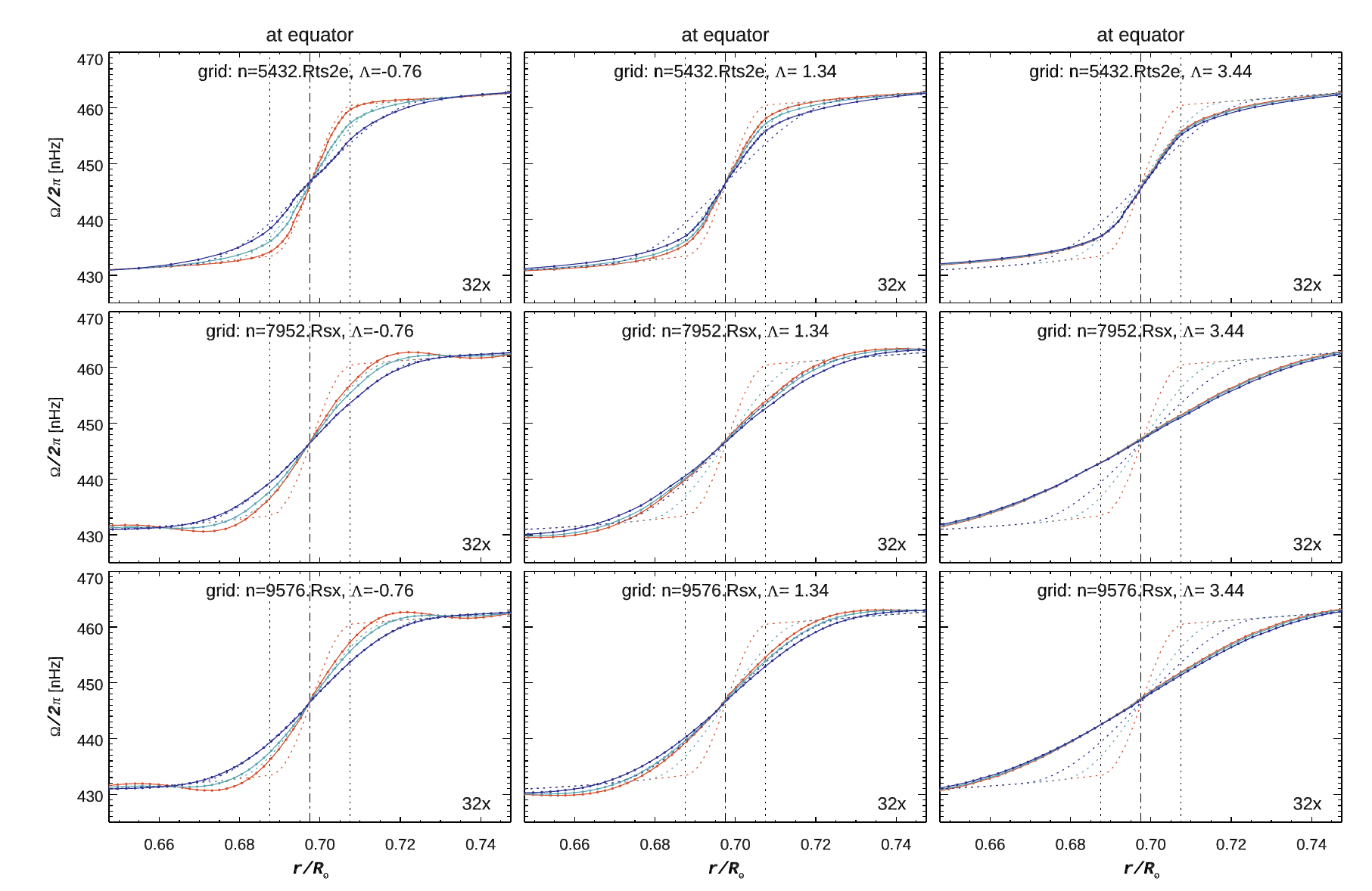} 
  \end{center}
  \caption{RLS solutions for noiseless simulations (solid lines with dots)
    superimposed over the true solutions (dashed lines) for the three models
    of the tachocline, shown using different colors (red for {\em narrow},
    green for {\em nominal}, and blue for {\em wide}).  Top to bottom panels
    correspond to different inversion grids ({\em aggressive}, {\em
      progressive I}, and {\em progressive II}), while left to right panels
    correspond to increasing smoothing (as per the trade-off coefficient
    $\Lambda$).}
  \label{fig:rls-noiseless}
\end{figure}

\begin{figure}
  \begin{center}
    \includegraphics[width=0.9\textwidth]{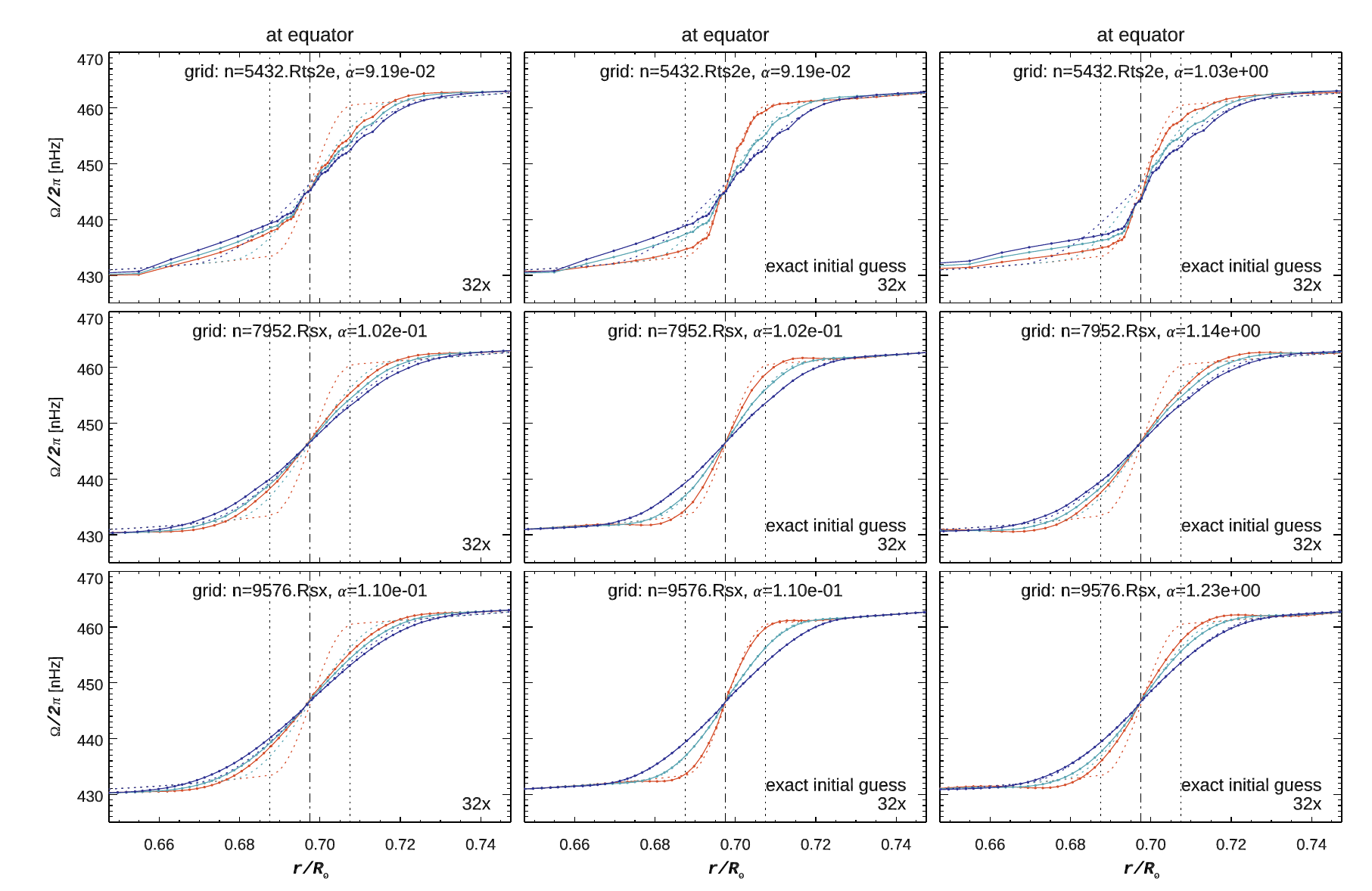} 
  \end{center}
  \caption{SART solutions for noiseless simulations (solid lines with dots)
    superimposed over the true solutions (dashed lines) for the three models
    of tachocline, shown using different colors (red for {\em narrow}, green
    for {\em nominal}, and blue for {\em wide}). Top to bottom panels
    correspond to different inversion grids ({\em aggressive}, {\em
      progressive I} and {\em progressive II}), while leftmost panel shows
    solutions when initial guess is set to a constant, while the middle and
    rightmost panels correspond to solutions when setting the initial to the
    true solutions and varying the smoothing factor $\alpha$ (\ie, smoothing
    increases with lower $\alpha$ values). The unique combination of the {\em
      progressive II} grid, exact initial solution and low smoothing (in the
    lowest middle panel) recovers almost perfectly the three models.}
  \label{fig:sart-noiseless}
\end{figure}

The effect of radial gridding on solutions derived from noiseless simulations
is shown in Fig.~\ref{fig:rls-noiseless} for the RLS method and in
Fig.~\ref{fig:sart-noiseless} for the SART method, whether using as an initial
guess a constant value, or the exact solution.

These figures show clearly how the RLS solution with a very high radial
resolution resolves the tachocline, when using a very low smoothing
trade-off. It also hints at a Gibbs-like phenomenon in the case of the narrow
tachocline.

For the SART method, the solutions computed using the very high radial
resolution display numerical noise that we trace to the rate of change of the
radial grid. Indeed, this noise is absent from the solutions for the the
progressive grids. Not surprisingly, the solutions for noiseless simulations,
when starting with the exact solution and when using very little smoothing,
recover the narrow tachocline. Yet adding more radial points to reach a lower
minimum resolution, using a progressive grid, recovers the solution even
better. And, not surprisingly either, adding some smoothing, in all cases,
leads to a set of solutions that no longer resolves the model with the most
narrow tachocline.

\begin{figure}
  \begin{center}
    \includegraphics[width=0.9\textwidth]{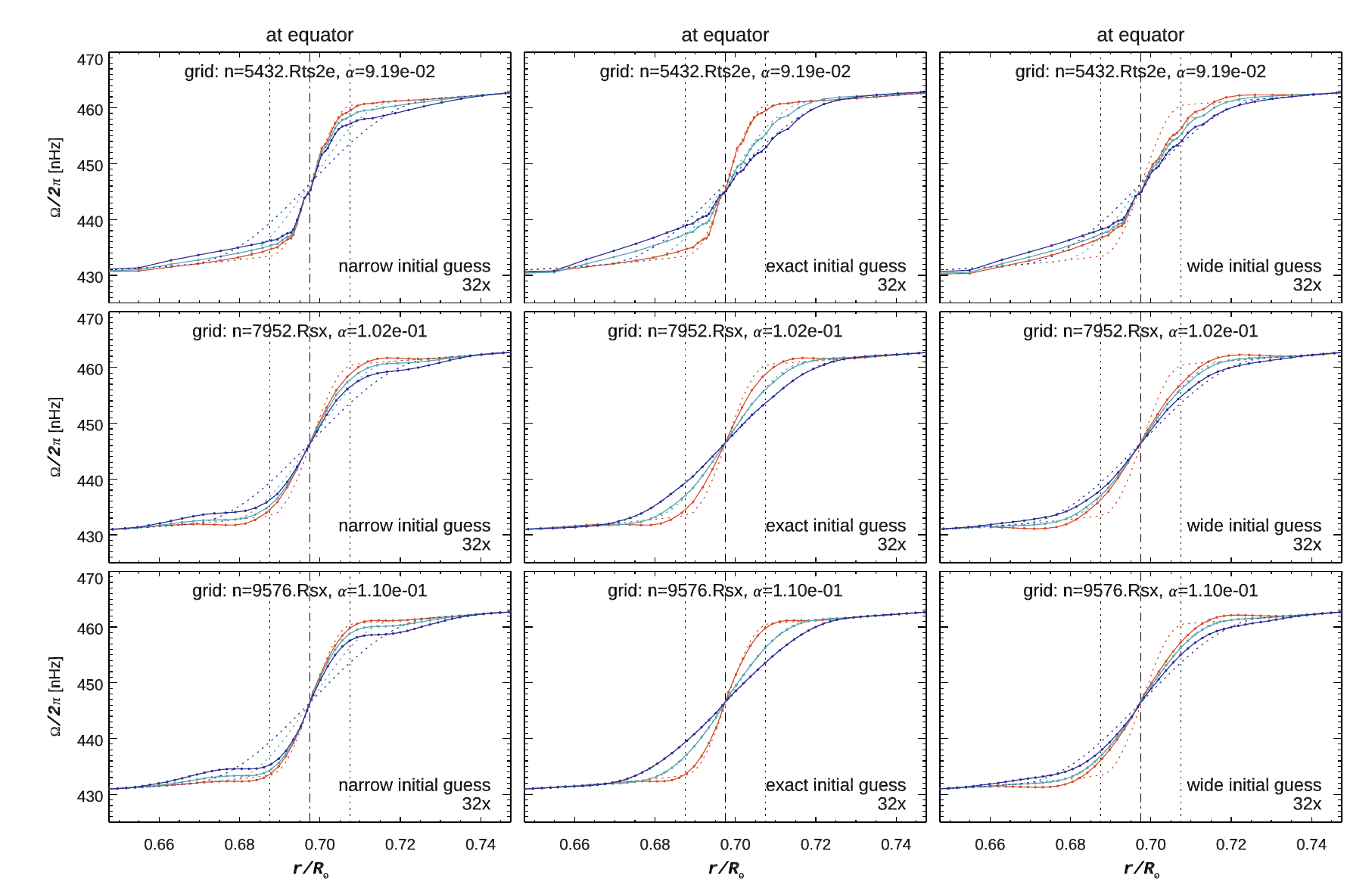} 
  \end{center}
  \caption{SART solutions for noiseless simulations (solid lines with dots)
    superimposed over the true solutions (dashed lines) for the three models
    (shown with different colors) and almost no smoothing. Top to bottom
    panels correspond to different inversion grids ({\em aggressive}, {\em
      progressive I} and {\em progressive II}), while left to right panels
    correspond to using different initial guesses (narrow, exact and wide
    respectively).}
  \label{fig:sart-initial-guess-almost-no-smoothing}
\end{figure}

\begin{figure}
  \begin{center}
    \includegraphics[width=0.9\textwidth]{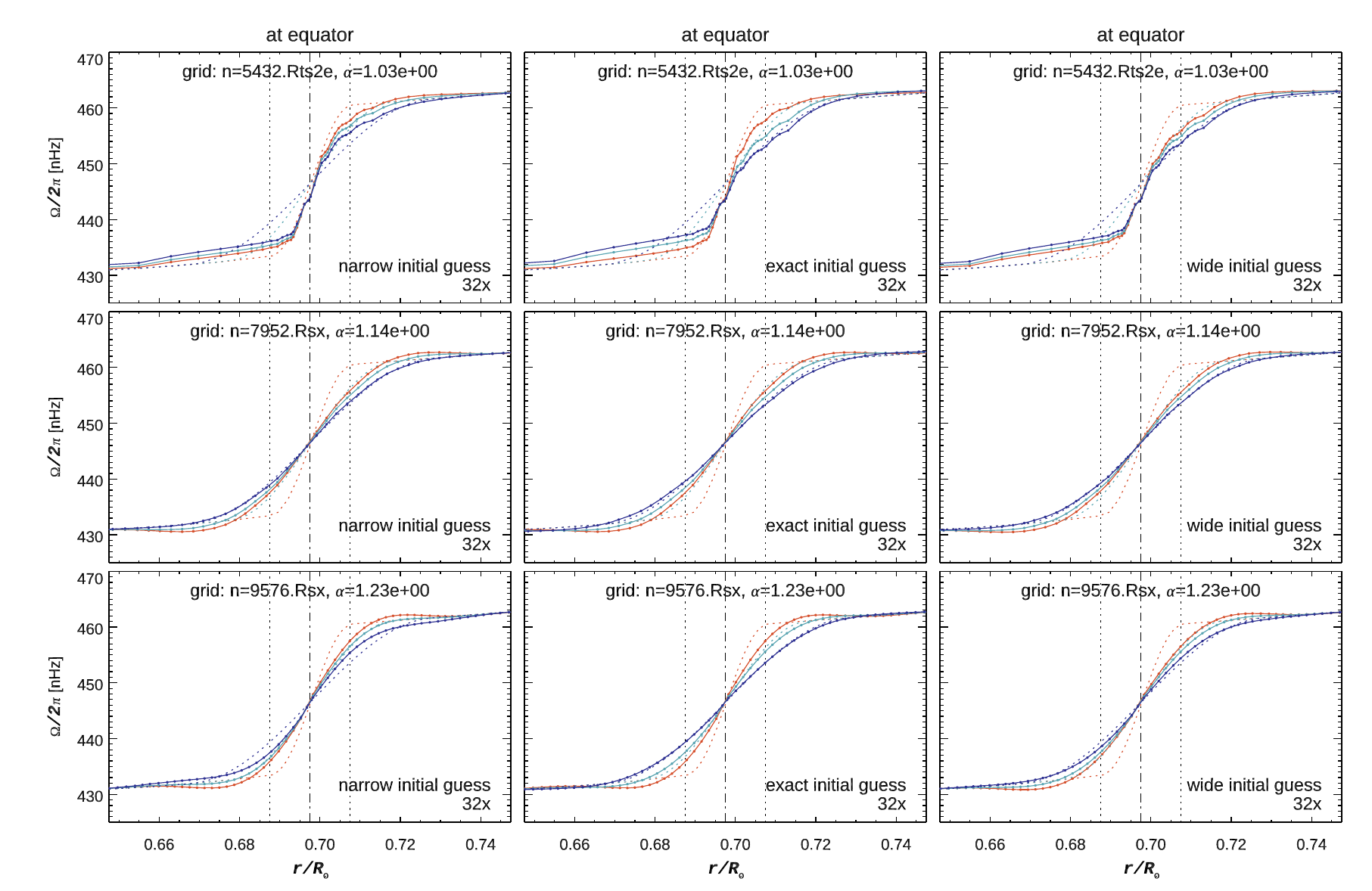} 
  \end{center}
  \caption{SART solutions for noiseless simulations (solid lines with dots)
    superimposed over the true solutions (dashed lines) for the three models
    (shown with different colors) and ``typical'' smoothing. Top to bottom
    panels correspond to different inversion grids ({\em aggressive}, {\em
      progressive I} and {\em progressive II}), while left to right panels
    correspond to using different initial guesses (narrow, exact and wide
    respectively).}
  \label{fig:sart-initial-guess-most-smoothing}
\end{figure}

We also need to check the effect of the choice of the initial guess, since
except for simulations, the exact solution is unknown. Henceforth we inverted
noiseless simulated data sets starting with a narrow, nominal or wide initial
guess for all 3 simulations (\ie, models with narrow, nominal, and wide
tachocline).
The resulting solutions when using almost no smoothing are presented in
Fig.~\ref{fig:sart-initial-guess-almost-no-smoothing}, while solutions when
using what we consider typical smoothing are shown in
Fig.~\ref{fig:sart-initial-guess-most-smoothing}. These plots indicate that in
the absence of noise and with almost no smoothing, the SART method with a too
narrow initial guess will not properly recover the wider tachoclines, yet this
becomes less of a problem when including typical smoothing.

\subsection{More Realistic Simulations}

\begin{figure}
  \begin{center}
    \includegraphics[width=0.9\textwidth]{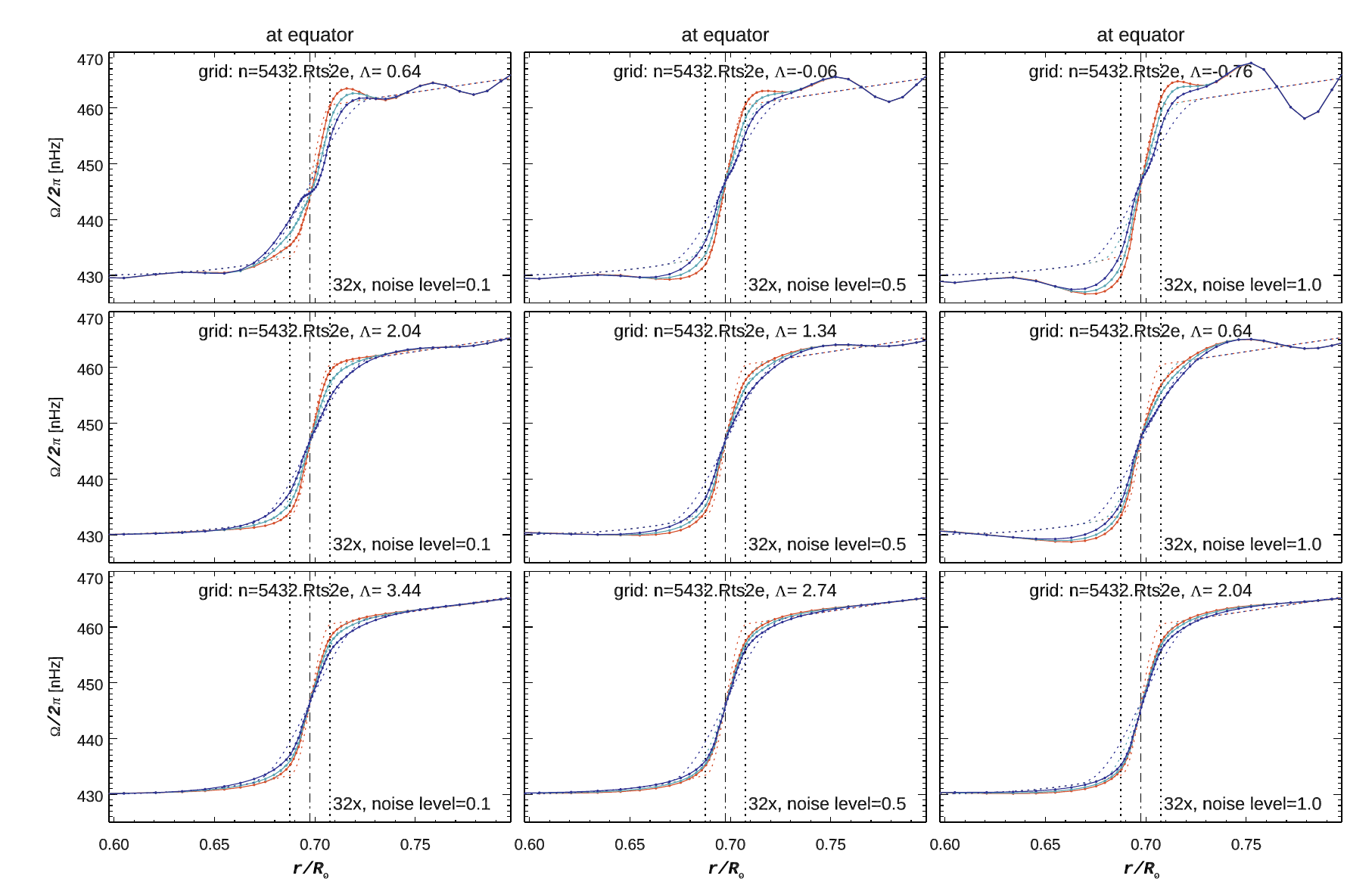} 
  \end{center}
  \caption{RLS solutions for noisy simulations (solid lines with dots)
    superimposed over the true solutions (dashed lines) for the three models
    (shown with different colors). Top to bottom panels correspond to
    increasing smoothing (as per the trade-off coefficient $\Lambda$), while
    left to right panels correspond to increasing levels of noise, with
    respect to the uncertainties resulting from fitting a $32 \times 72$d time
    series.}
  \label{fig:rls-noise}
\end{figure}

\begin{figure}
  \begin{center}
    \includegraphics[width=0.9\textwidth]{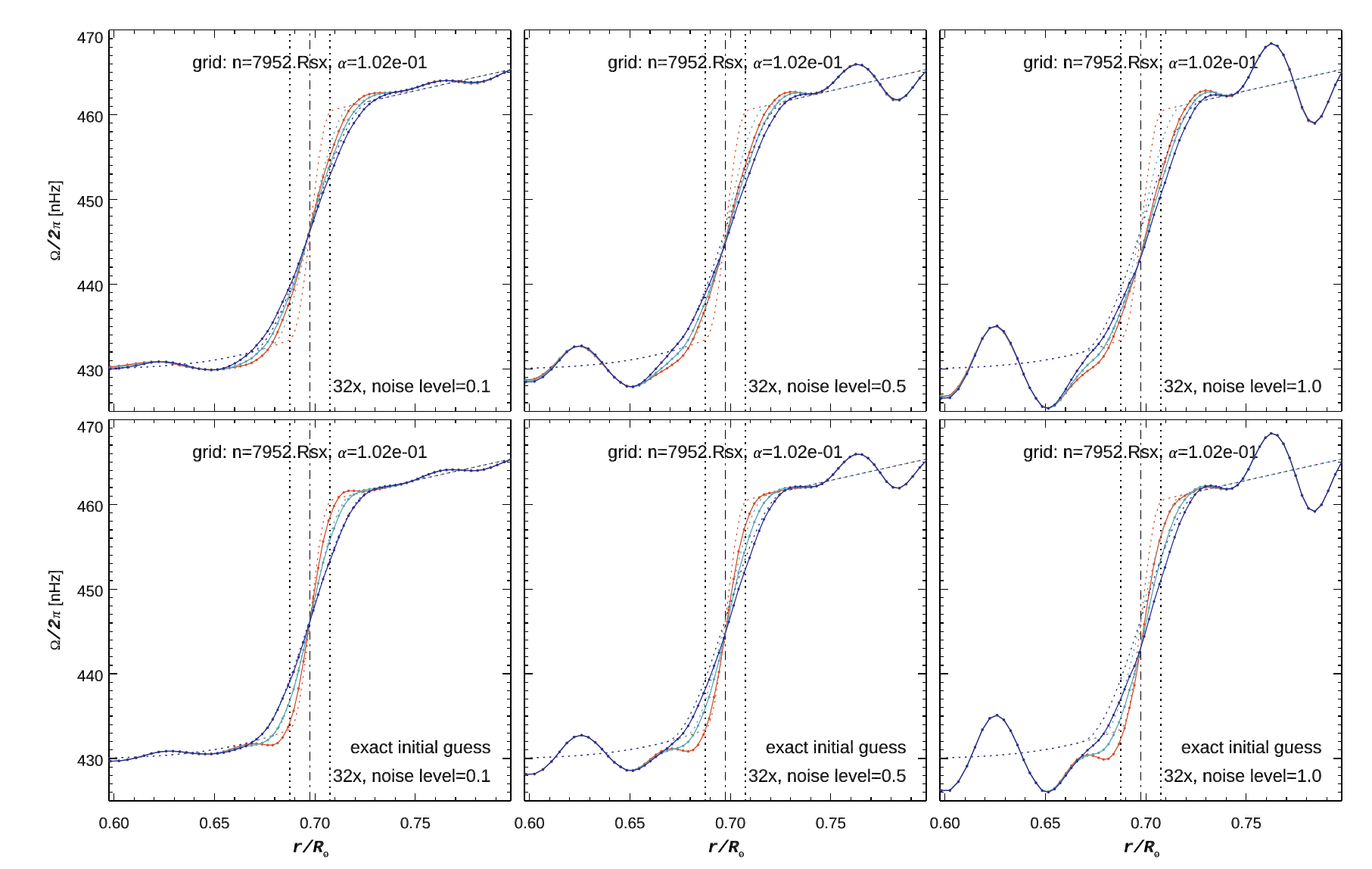}\\
    \includegraphics[width=0.9\textwidth]{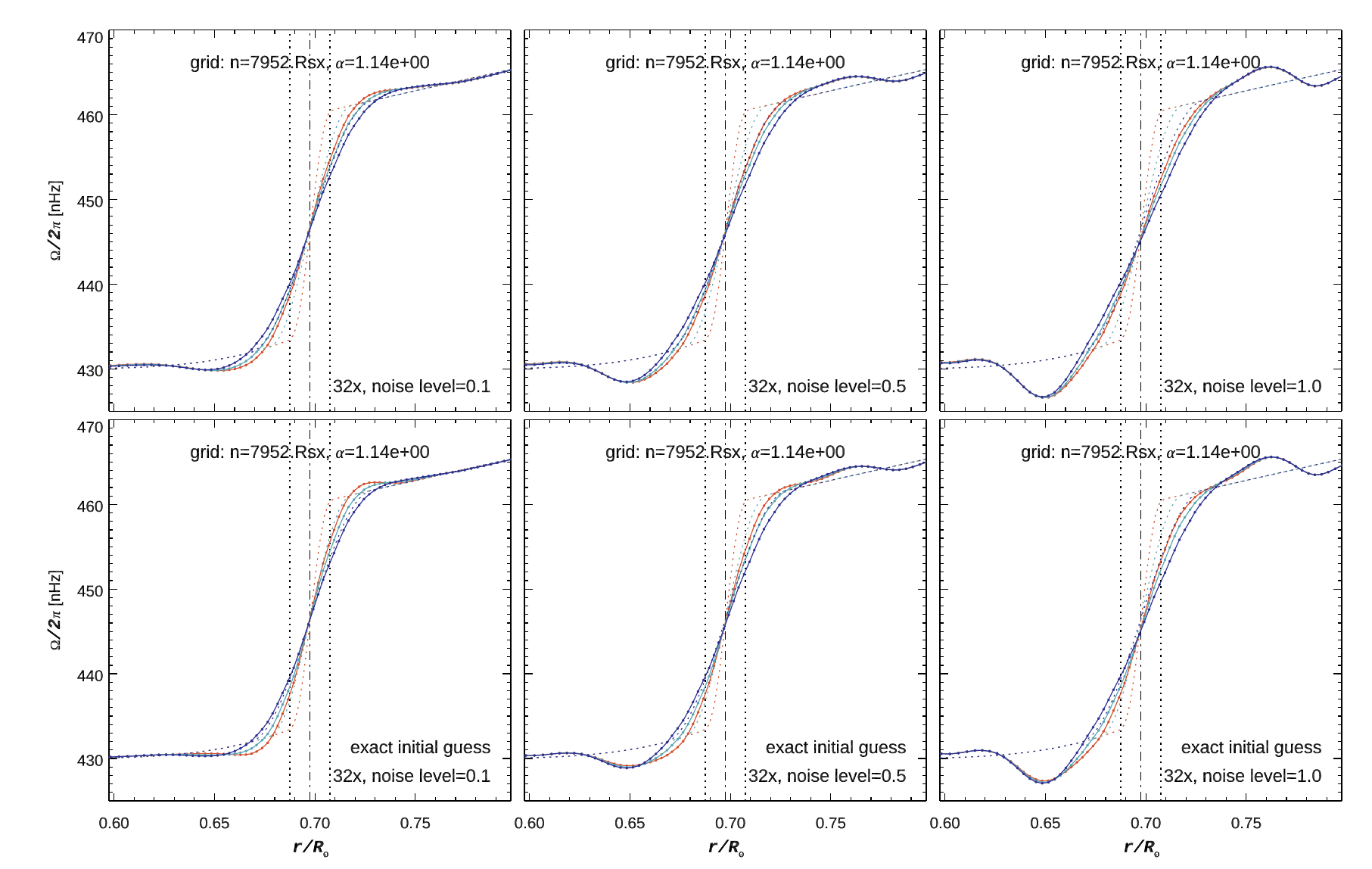} 
  \end{center}
  \caption{SART solutions for noisy simulations (solid lines with dots)
    superimposed over the true solutions (dashed lines) for the three models
    (shown with different colors), using the {\em progressive I} grid. The top
    two rows correspond to almost no smoothing, while the bottom two rows to
    ``typical'' smoothing (as per value of $\alpha$). Rows alternate cases
    with constant or exact initial guess, while columns correspond to
    increasing levels of noise, with respect to the uncertainties resulting
    from fitting a $32 \times 72$d time series (from left to right).}
  \label{fig:sart-noise}
\end{figure}

Since in practice actual observations have some level of noise, it is a lot
more instructive to invert simulations with random noise. Solutions for noisy
simulations using the RLS method are presented in Fig.~\ref{fig:rls-noise} and
when using the SART method, and two levels of smoothing, in
Fig.~\ref{fig:sart-noise}.

What these figures show is that in the presence of noise, the RLS and SART
solutions display a strong Gibbs-like effect. While RLS solutions using the
so-called aggressive grid looked great when using noiseless splittings, adding
any level of noise results in large excursions that gets damped with
additional smoothing.

The challenge when using actual observations remains the optimal and objective
selection of the trade-off between smoothing and noise amplification. Yet,
with these caveats in mind, using the aggressive radial grid remains a viable
strategy.  In contrast, the SART method behaves a lot better when using the
progressive grid, but remains affected by Gibbs-like excursions without
appropriate smoothing. Yet, using an informed initial guess, while tricky and
misleading in noiseless cases, offers limited improvement in the noisy cases.

\section{Results from Observations}\label{S-R-Observations}

From the tables of individual modes, $\nu_{n,\ell,m}$, we compute the scaled
rotational splittings, $s_{n,\ell,m}$, and perform an outliers rejection. The
scaled rotational splittings are defined as
\begin{equation}
    s_{n,\ell,m}=\frac{\delta\nu_{n,\ell,m}}{2 m}
\end{equation}
The uncertainties of the scaled splittings, $\sigma_{s_{n,\ell,m}}$, are
derived from the individual modes uncertainties.

The outliers rejection consists in fitting a straight line to the scaled
splittings, $s_{n,\ell,m}$, with respect to the ratio ${m}/{\ell}$, using a
3-$\sigma$ rejection when performing this fit. Scaled splittings that depart
from that straight line by some value, $r_s$, are rejected, \ie, when
\begin{equation}
    |s_{n,\ell,m}-f(\frac{m}{\ell})| > r_s
\end{equation}
as well as values whose reduced differences magnitude, \ie, the difference
divided by the uncertainty, exceed some threshold, $t_s$, \ie, when
\begin{equation}
    \left|\frac{s_{n,\ell,m}-f(\frac{m}{\ell})}{\sigma_{s_{n,\ell,m}}}\right| > t_s
\end{equation}
We tested several combinations of these two thresholds and settled on
$r_s=200$ nHz and $t_s=1.5$ as a good trade-off between limiting the amount of
rejection and yet eliminating most of the outliers.

\begin{figure}
  \begin{center}
    \includegraphics[width=.9\textwidth]{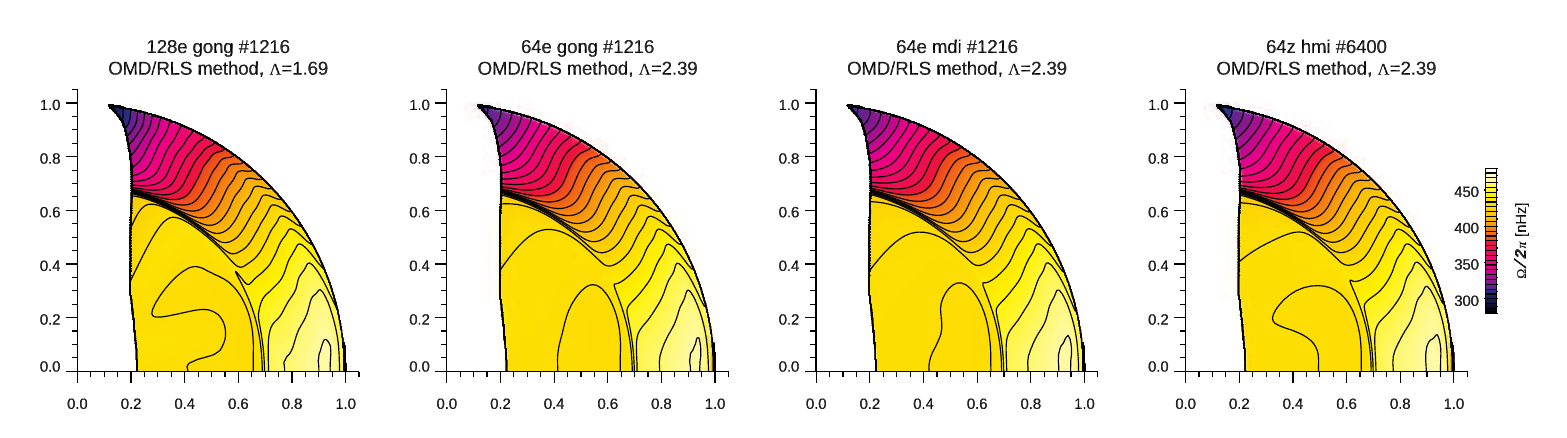} 
    \includegraphics[width=.9\textwidth]{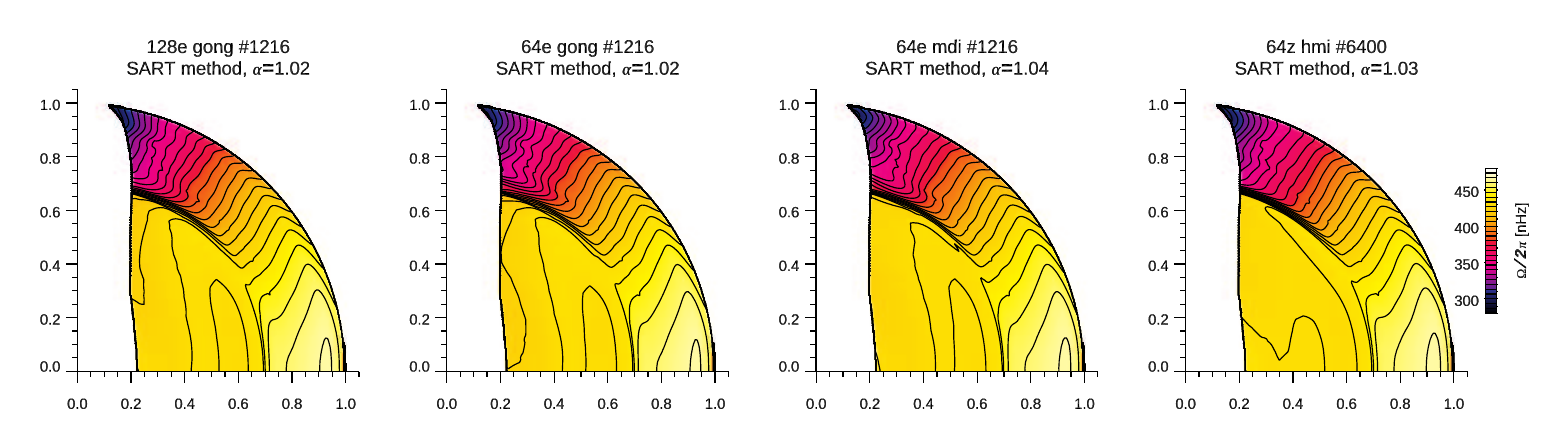} 
  \end{center}
  \caption{Inverted rotation rate derived from splittings resulting from
    fitting very long time series, i.e., 25.2 years of GONG observations
    (leftmost panel) and long time series, i.e., 12.6 years of GONG, MDI or
    HMI observations (rightmost 3 columns), using either RLS or SART inversion
    methodology (top versus bottom row).}
  \label{fig:obs-2d}
\end{figure}

The resulting rotational rates, for four data sets and both inversion methods,
are shown in an overall two-dimensional representation in Cartesian
coordinates in Fig.~\ref{fig:obs-2d}. The main features of the resulting
rotation rate are the well-established near solid body rotation in the
radiative interior, the differential rotation nearly constant on cones in the
convection zone, a narrow tachocline near the base of the convection zone and
a near-surface shear layer.

\subsection{Results Using Very Long Time Series}

\begin{figure}
  \begin{center}
    \includegraphics[width=.9\textwidth]{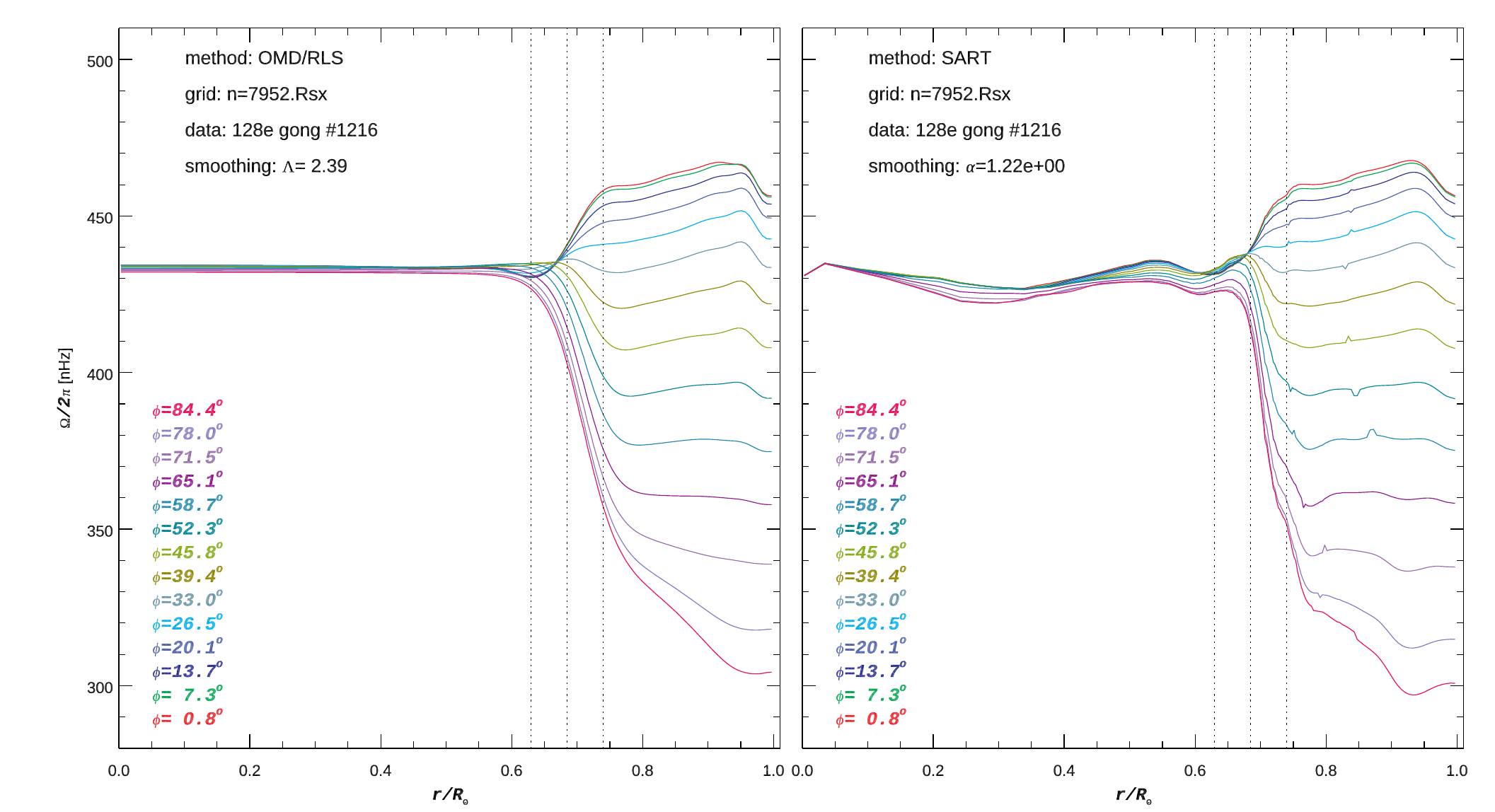} 
   \end{center}
  \caption{Inverted rotation rate derived from splittings resulting from
    fitting a very long time series, i.e., 25.2 years of GONG observations,
    plotted as a function of radius for a set of latitudes, using either the
    RLS or SART inversion methodology (left versus right panels).}
  \label{fig:obs-128e-cuts}
\end{figure}

A different representation of the inverted rotation rate derived from fitting
a very long time series is shown in Fig.~\ref{fig:obs-128e-cuts}, when using
in both cases the lower resolution progressive grid. The tachocline and the
near-surface shear are clearly visible.

In order to characterize the tachocline, we followed the methodology used in
\citet{BasuEtal-2024} where the tachocline is parametrized as a
sigmoid. Hence, at each latitude, we fit a sigmoid, complemented by two slopes
and a constant, to the inverted profiles as a function of radius, namely
\begin{equation}
  \tilde{\Omega}(r,\phi) =
  \Omega_{o,\phi}
    + \frac{\Delta\Omega_\phi}{1+e^{\frac{r-r_{d,\phi}}{w_{d,\phi}}}} +
    \left\{ \begin{array}{lr}
         s_{-,\phi}\,(r-r_{d,\phi}) & \mbox{~~~~~ for $r < r_{d,\phi}$} \\
         s_{+,\phi}\,(r-r_{d,\phi}) & \mbox{~~~~~ for $r > r_{d,\phi}$} \end{array} \right.
\end{equation}
where $\tilde{\Omega}$ is the inverted rotation rate, $r$ the radius, $\phi =
\pi/2 - \theta$ the latitude, $ r_{d,\phi}$ the location of the sigmoid
discontinuity, $w_{d,\phi}$ the width of the sigmoid, $\Delta\Omega_\phi$ the
amplitude of the sigmoid jump, $\Omega_{o,\phi} + \frac{\Delta\Omega_\phi}{2}$
is the rotation at $r=r_d$ and ($s_{-,\phi}$, $s_{+,\phi}$) the slopes below
or above $r_{d,\phi}$.

We use the limited radial range $0.43 < r/R_{\odot} < 0.97$ to fit this model,
and note that by including slopes, the actual amplitude of the tachocline jump
is given, at each latitude, by
\begin{equation}
  \Delta\Omega_{J,\phi} = \Delta\Omega_{\Phi}\,(\frac{e^{F_\epsilon}-1}{e^{F_\epsilon}+1})
  + (s_{+,\phi} + s_{-,\phi})\, F_\epsilon\,w_{d,\phi} 
\end{equation}
where $F_\epsilon = \log(\frac{1}{\epsilon}-1)$. We chose to evaluate this for
$\epsilon=0.5\times10^{-3}$, namely the range for which the sigmoid goes from
$\epsilon$ to $1-\epsilon$, or 0.999 of its variation, hence $F_\epsilon =
7.6$.

\subsection{Tachocline Characteristics}

\begin{figure}
  \begin{center}
    \includegraphics[trim=20 0 0 0,clip,width=.25\textwidth]{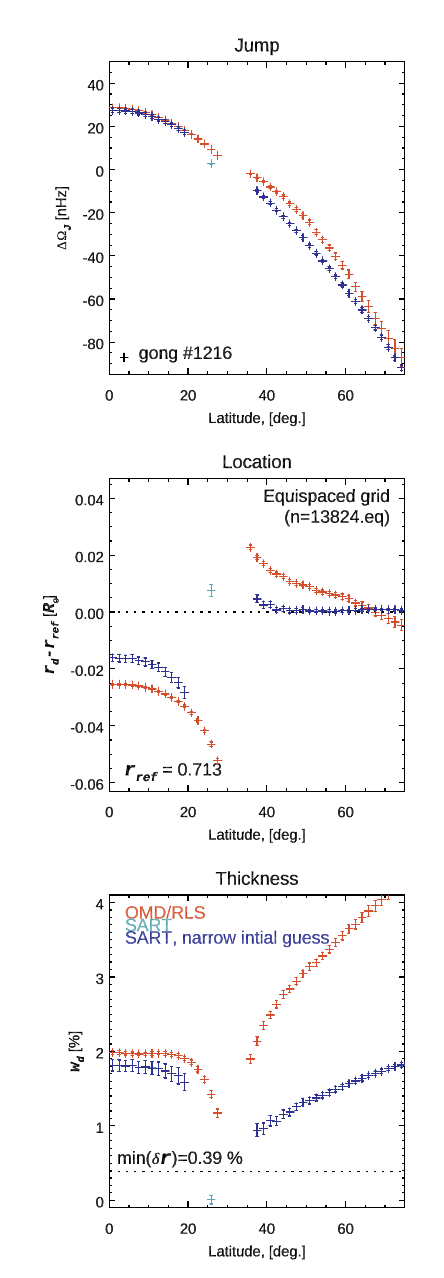}
    \includegraphics[trim=20 0 0 0,clip,width=.25\textwidth]{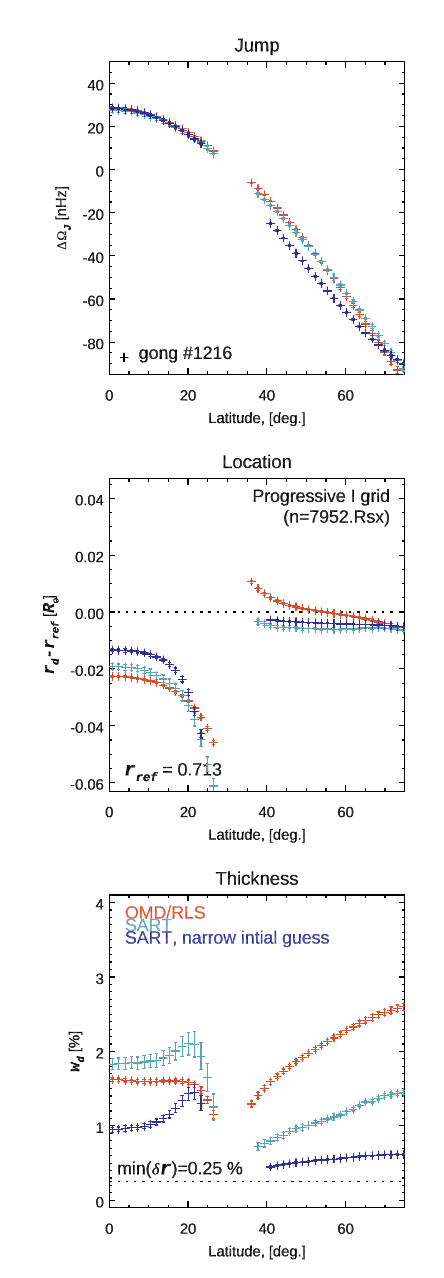}
    \includegraphics[trim=20 0 0 0,clip,width=.25\textwidth]{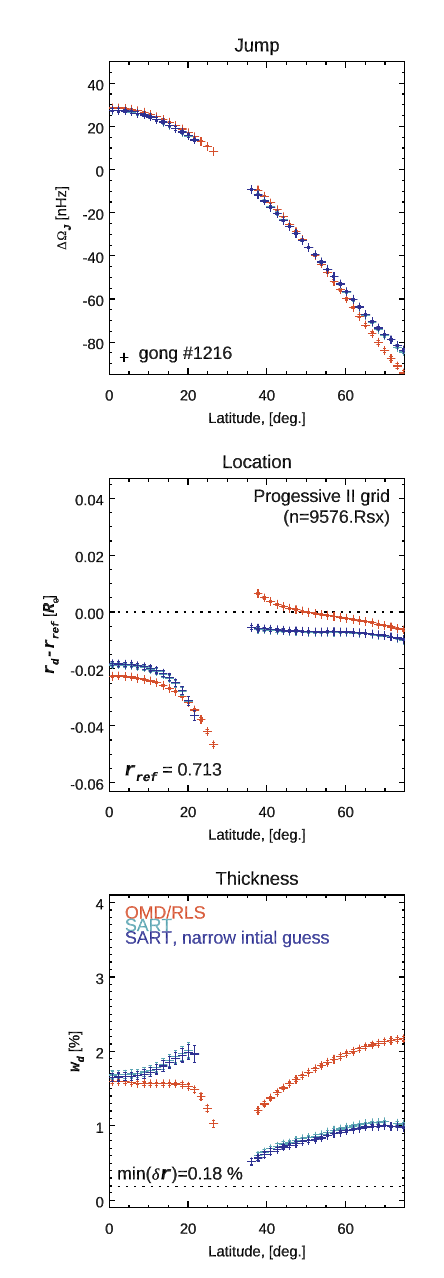}
    \includegraphics[trim=20 0 0 0,clip,width=.25\textwidth]{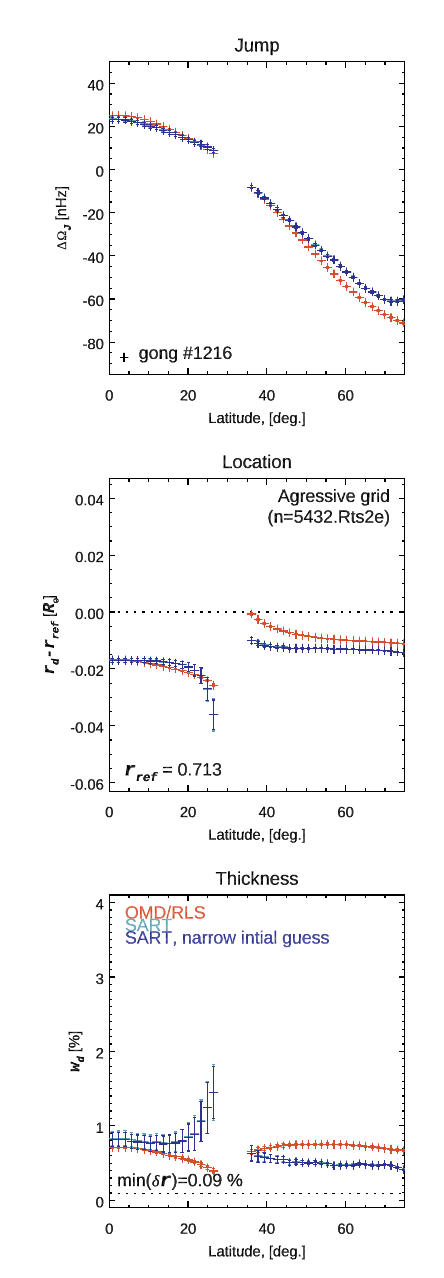} 
  \end{center}
  \caption{Characteristics of the tachocline resulting from fitting, around
    the tachocline, a sigmoid with two slopes to inverted rotation rates
    derived from splittings resulting from fitting a very long time series,
    i.e., 25.2 years of GONG observations. Namely, from top to bottom, the
    total jump, $\Delta\Omega_J$, location, $r_d$, and thickness, $w_d$, are
    plotted as a function of latitude, for inversions using the RLS and SART
    methods for, from left to right, four different inversion grids, \ie, an
    equispaced, two progressive and one aggressive radial density grid.  The
    location is shown with respect to the base of the convection zone, \ie,
    $r=0.713R_{\odot}$, while, for comparison with the inferred widths, the
    resolution of the inversion grid at the tachocline is indicated by the
    horizontal dotted lines.}
  \label{fig:sigmoid-results-128e}
\end{figure}

The parameters resulting from fitting our sigmoid model to inverted rotations
inferred from splittings determined by fitting the 25.2 year long GONG time
series are shown in Fig.~\ref{fig:sigmoid-results-128e}. The total jump,
$\Delta\Omega_J$, the location, $r_d$, and the thickness, $w_d$, are plotted
as a function of latitude, when using either inversion method, or when using
the SART inversion with a narrow initial guess, and for four different
inversion grids.
Not surprisingly the sigmoid model fitting fails at latitudes where the
amplitude of the jump is marginal, since the jump changes sign between low and
high latitudes (\ie, around $\phi=32$\textdegree).

These estimates of the total jump around the tachocline, $\Delta\Omega_J$,
are, up to 60\textdegree, consistent whatever method or grid is being used,
except for the aggressive grid at high latitudes.
The estimates of the location of the tachocline, $r_d$, are also overall
consistent with respect to the inversion methodology or inversion grid, with
some notable exceptions. For all cases, one salient feature is a discontinuity
of that location between values derived at low and high latitudes. That
discontinuity is more pronounced for the RLS solutions and the precise
location at the higher latitudes is somewhat grid dependent.
As for estimates of the thickness of the tachocline, $w_d$, the figures show
some inconsistencies between methodology (\ie, the RLS returning
systematically a wider tachocline), but more significantly the estimated width
gets smaller as the resolution of the grid around the tachocline gets smaller.
Hence, despite its pitfalls, inversions that use the high resolution
aggressive grid infer that the tachocline is quite narrow, namely with a
FWHJ\footnote{The full width for half jump, \ie, the radial extent of 50\% of
  the jump, or from 25\% to 75\%, is $2.2\,w_d$, or $2\,F_\epsilon\,w_d$ with
  $F_\epsilon$ estimated for $\epsilon=0.25$.}  below 1\%.
These consistencies and inconsistencies (\ie, inferences from different
methods and different grids) are consistent with the results of our
simulations.

\subsection{Effective Location and Averaging Kernels}

A possible concern on the robustness of the derived characteristics of the
tacho\-cline using the fitting of a sigmoid model, as shown in
Fig.~\ref{fig:sigmoid-results-128e}, is the issue of ``{\em effective
  location}.'' Indeed the estimate of the rotation rate inferred by inversions
is the convolution of the underlying {\em true} solution by the corresponding
averaging kernels, not the true value at the target location. Therefore we
checked if this convolution is not displacing in a significant way the
effective location of the inferred rate from the target location.

In order to assess this we need to define this effective location and decided
to use the center of gravity of the averaging kernels. To get an accurate
measure of the center of gravity, we computed these averaging kernels on a
much higher density grid than the inversion grid, yet not as high as the grid
used to compute the rotational kernels, since using such a very high
resolution grid would lead to matrices too large to handle. We tested three
resolutions to confirm that our resolution was good enough.

We not only need to properly resolve these averaging kernels, but we need to
handle the truncation of the kernels to one hemisphere. Indeed, let us
consider the cases of the averaging kernels computed for target locations
nearly at the equator.\footnote{Since we use piecewise constant functions for
  the discretization of the inverse problem, our lowest latitude, being
  effectively at the midpoint, will be half the latitude increment.}  If we
compute the center of gravity of $K^{(a)}(r,\theta)$ for $0 \le \theta \le
\frac{\pi}{2}$ the resulting latitude of the center of gravity will be some
positive relatively large value, yet because of the hemispherical symmetry of
the problem, the averaging kernel of the symmetric solution is that
symmetrized kernel. This skewness will, of course, extend to a range of low
latitudes.

\begin{figure}
  \begin{center} %
    \includegraphics[width=.425\textwidth]{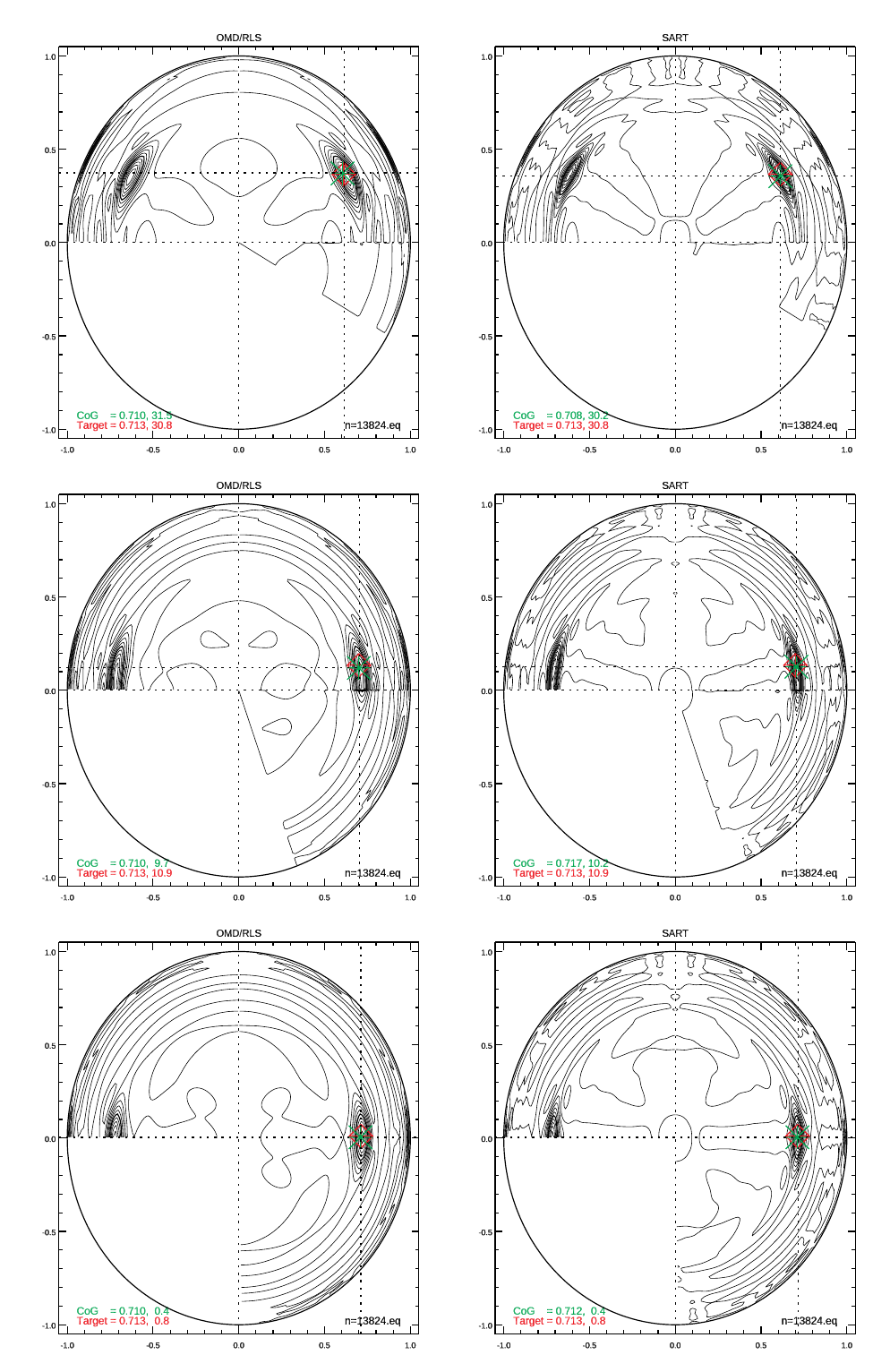} 
    \includegraphics[width=.425\textwidth]{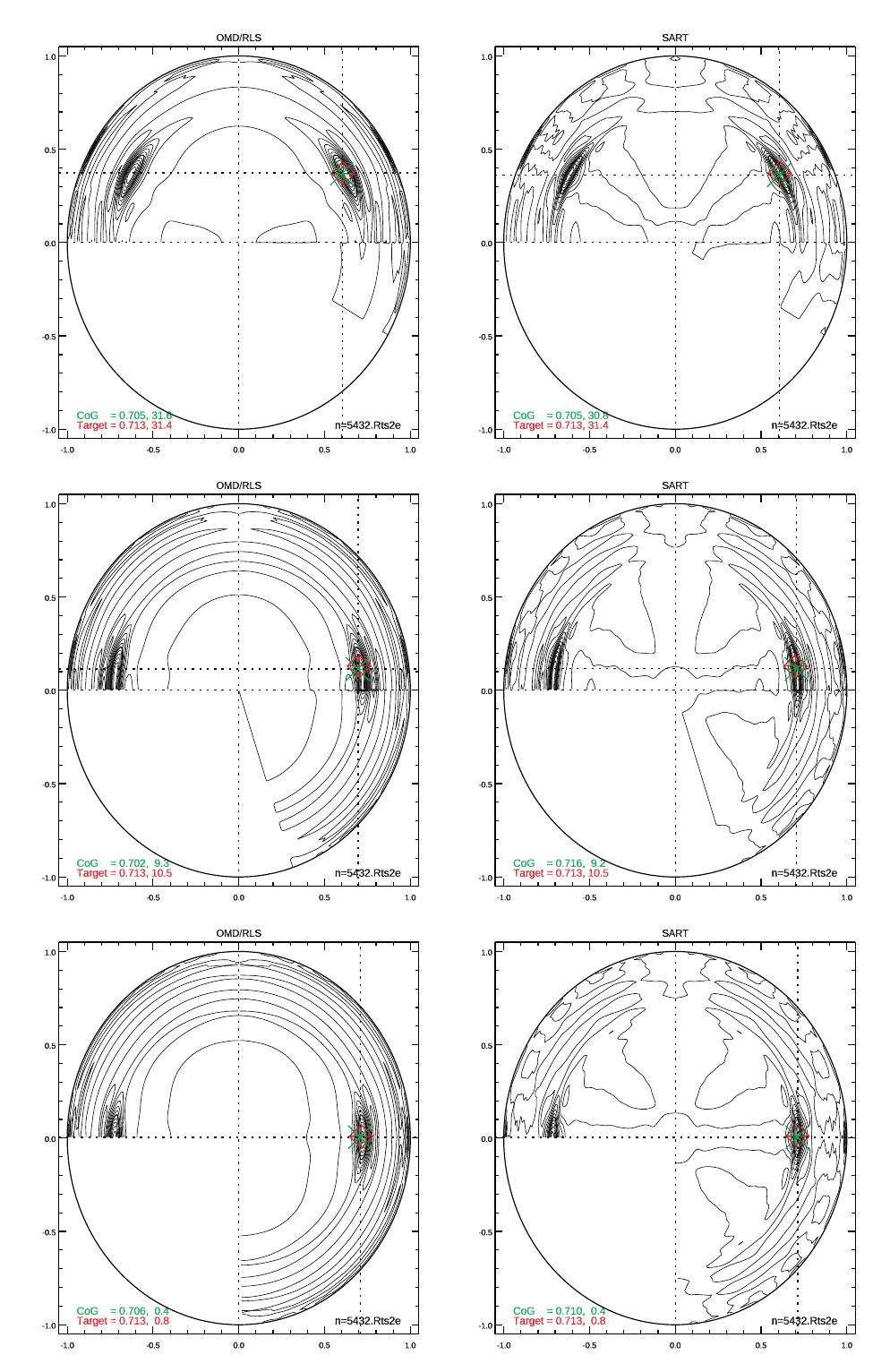} 
  \end{center}
  \caption{Illustration of the extended averaging kernels, for three latitudes
    (top to bottom) at $r/R_{\odot}=0.714$, for 
    two grids (equispaced and aggressive grids, columns)
    and both methods (even/odd columns). The averaging kernel is shown on the
    left, while the extended averaging kernels is shown on the right, with the
    50\% contour in bold. The target location is indicated in red while the
    center of gravity of the extended kernel is indicated in green.}
  \label{fig:extended-kernels}
\end{figure}

In order to mitigate this, we define and compute what we call extended
averaging kernels, that cover the full $-\frac{\pi}{2}\le\theta \le
\frac{\pi}{2}$ range, and compute the center of gravity of this extended
averaging kernel. This extension consists in assigning the values obtained at
latitudes above the target latitude to values at negative latitudes, separated
by the same distance from the target latitude, namely, for positive $\theta$
we use
\begin{equation}
    K^{(a_{\rm e})}_{i,j}(r,-\theta) =  K^{(a)}_{i,j}(r,\theta+(\theta+\theta_j))
\end{equation} 
for $0 \le 2\,\theta+\theta_j \le \frac{\pi}{2}$.
Examples of extended averaging kernels are illustrated in
Fig.~\ref{fig:extended-kernels} for three latitudes, one radius and three
inversions grids.

\begin{figure}
  \begin{center}
    \fbox{\href{https://drive.google.com/file/d/1dNQy6FAZrowGVJS9Mvj0dVz8PxgypsLQ/view}%
               {\includegraphics[width=.45\textwidth]{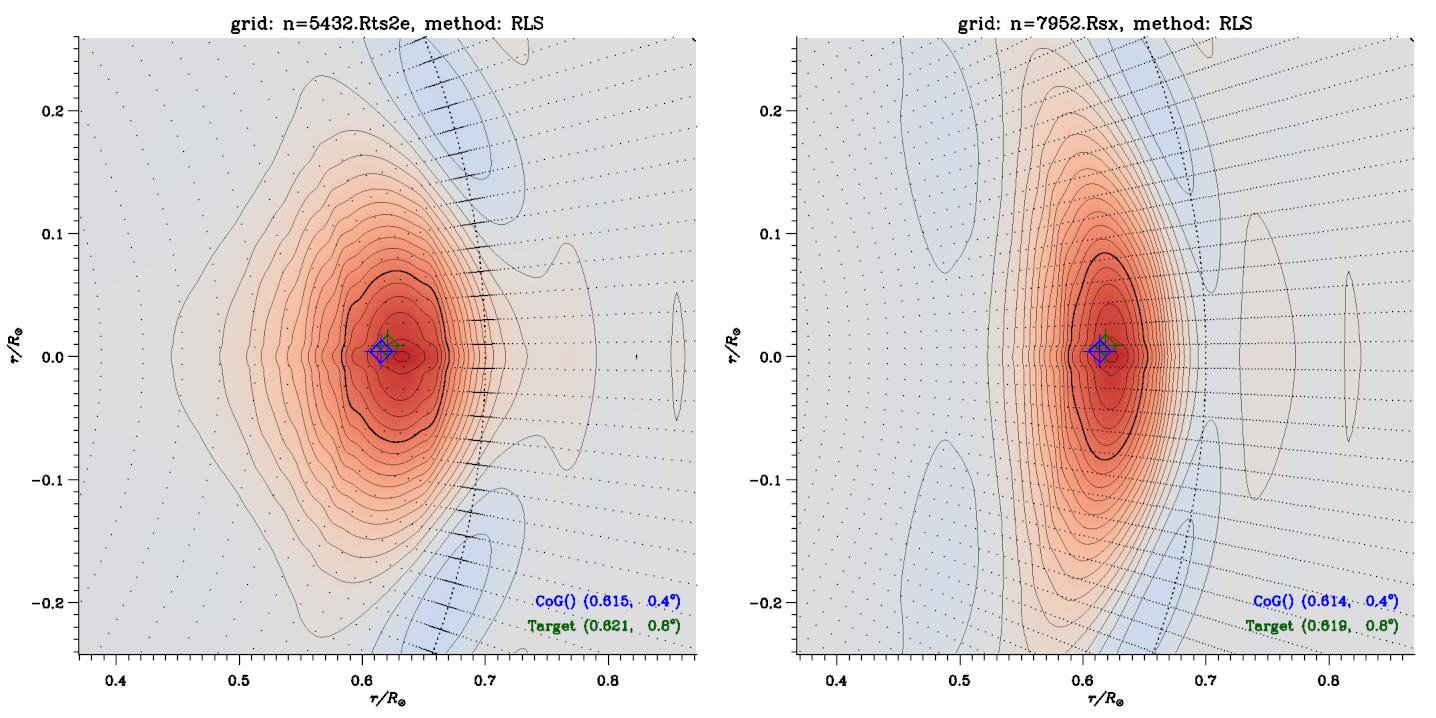}}} 
    \fbox{\href{https://drive.google.com/file/d/1QlLjHpEIvc-zlBQU08DrDpTuKwYYyXob/view}%
               {\includegraphics[width=.45\textwidth]{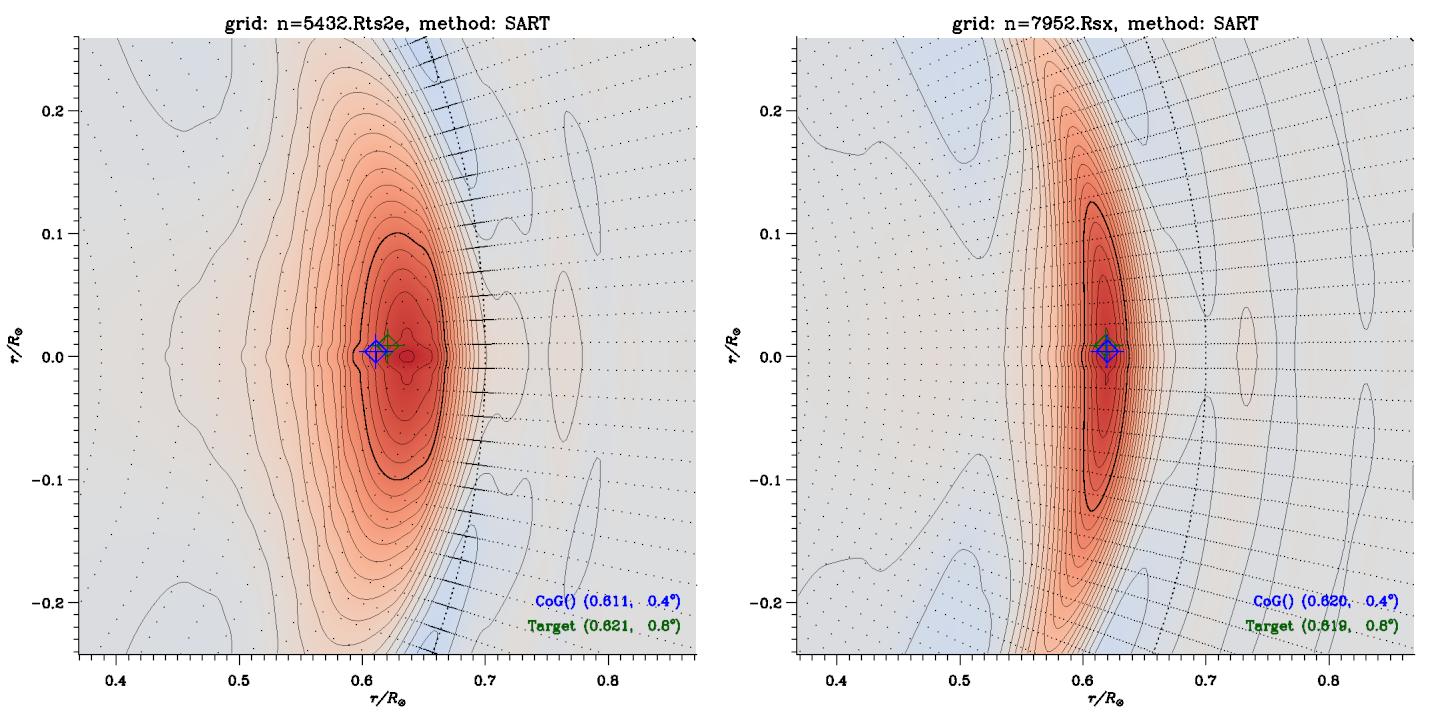}}} \\ 
    {\tiny ~~~~(a) $\phi=0.0^{\rm o}$ RLS \hspace*{.30\textwidth} (b) SART \hfill ~} \\
    \fbox{\href{https://drive.google.com/file/d/1NFg2kqM34Sd_Wv_ZRLtmjlJmXE1W01zr/view?usp=drive_link}%
               {\includegraphics[width=.45\textwidth]{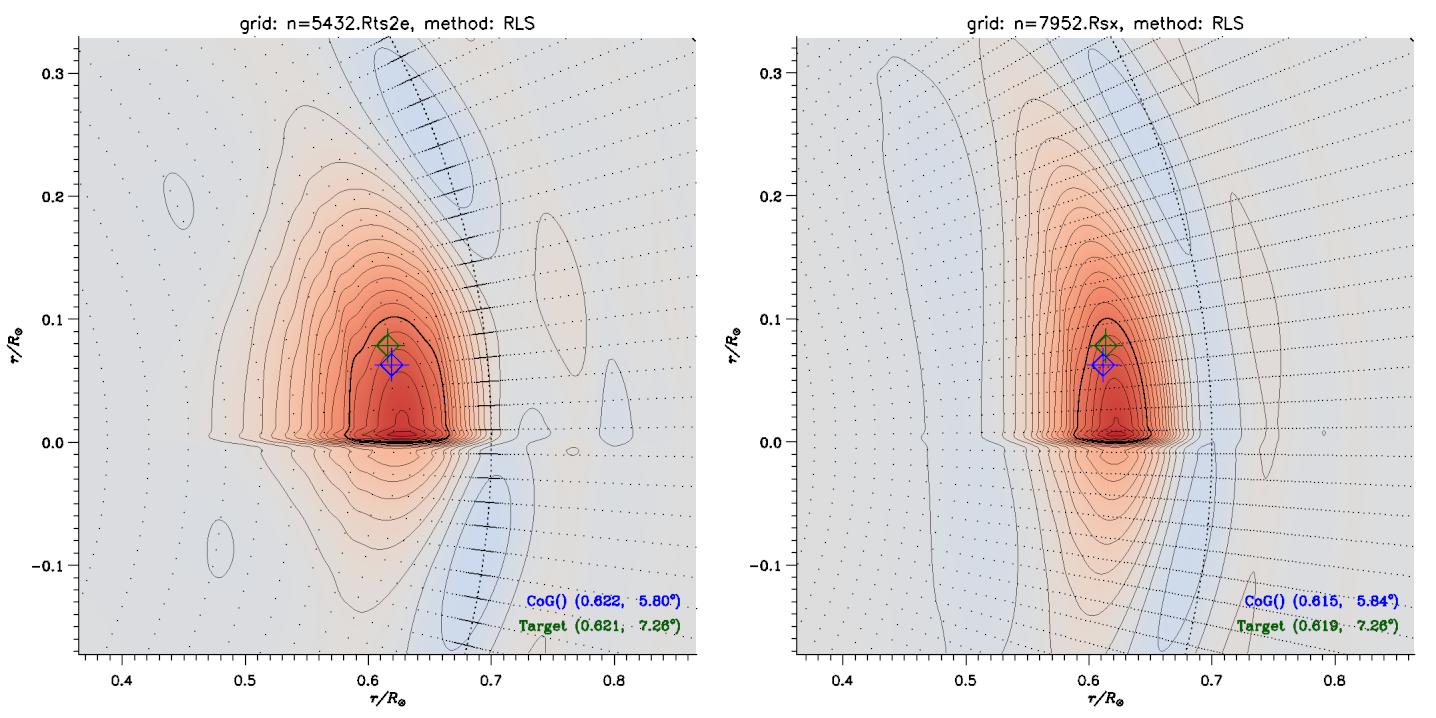}}} 
    \fbox{\href{https://drive.google.com/file/d/1bTDHL7nalyY75G0Qh3Waf29GnMo1YzTR/view?usp=drive_link}%
               {\includegraphics[width=.45\textwidth]{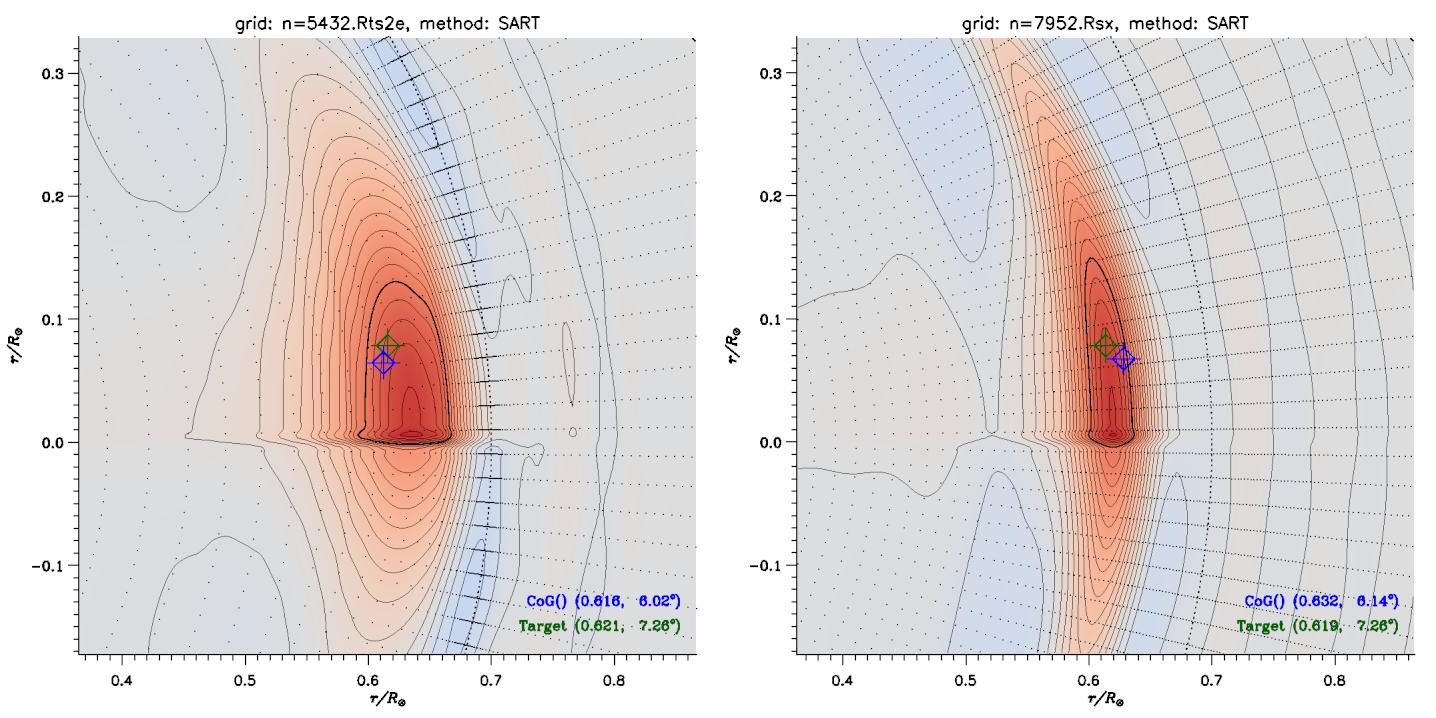}}} \\ 
    {\tiny ~~~~(c) $\phi=7.5^{\rm o}$ RLS \hspace*{.30\textwidth} (d) SART \hfill ~} \\
    \fbox{\href{https://drive.google.com/file/d/1kCyzHP9WAge0mnLtzZtPF_tiNBik21Bq/view?usp=drive_link}%
               {\includegraphics[width=.45\textwidth]{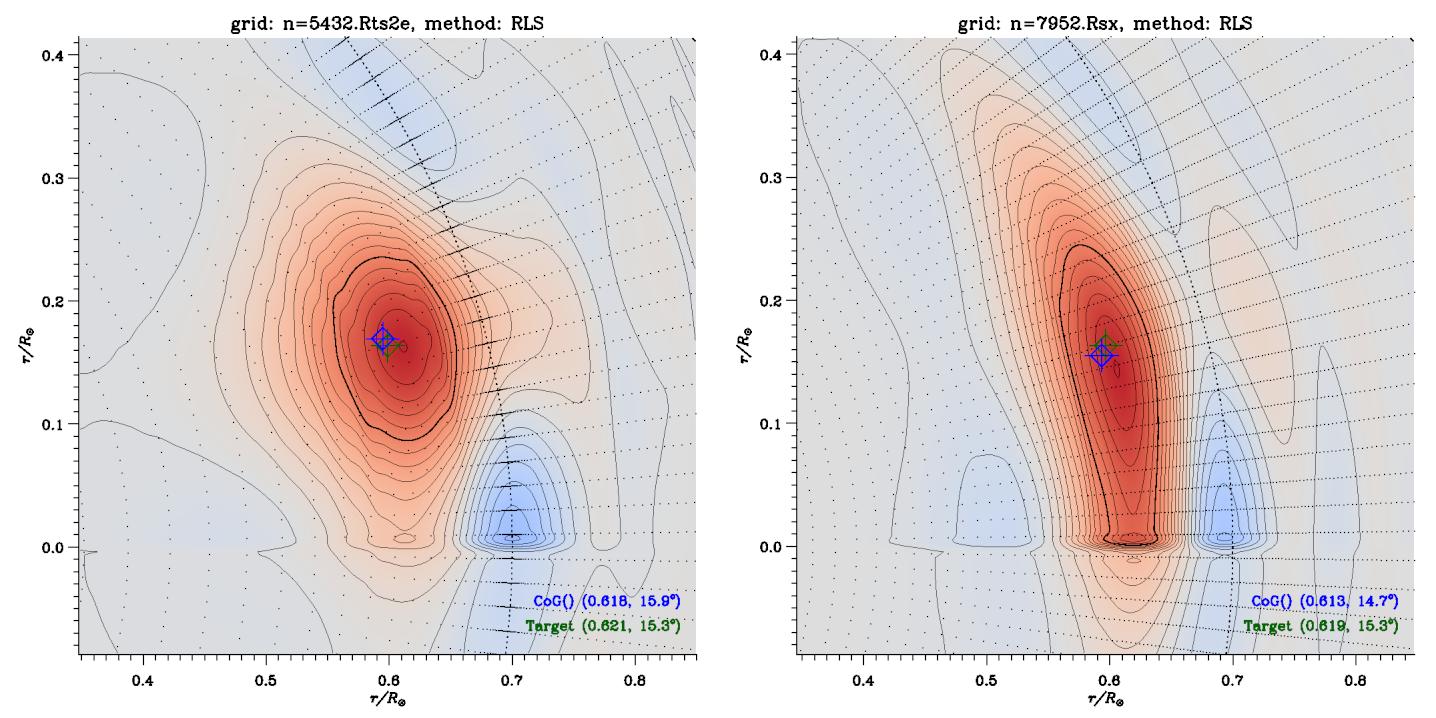}}} 
    \fbox{\href{https://drive.google.com/file/d/1kCyzHP9WAge0mnLtzZtPF_tiNBik21Bq/view?usp=drive_link}%
               {\includegraphics[width=.45\textwidth]{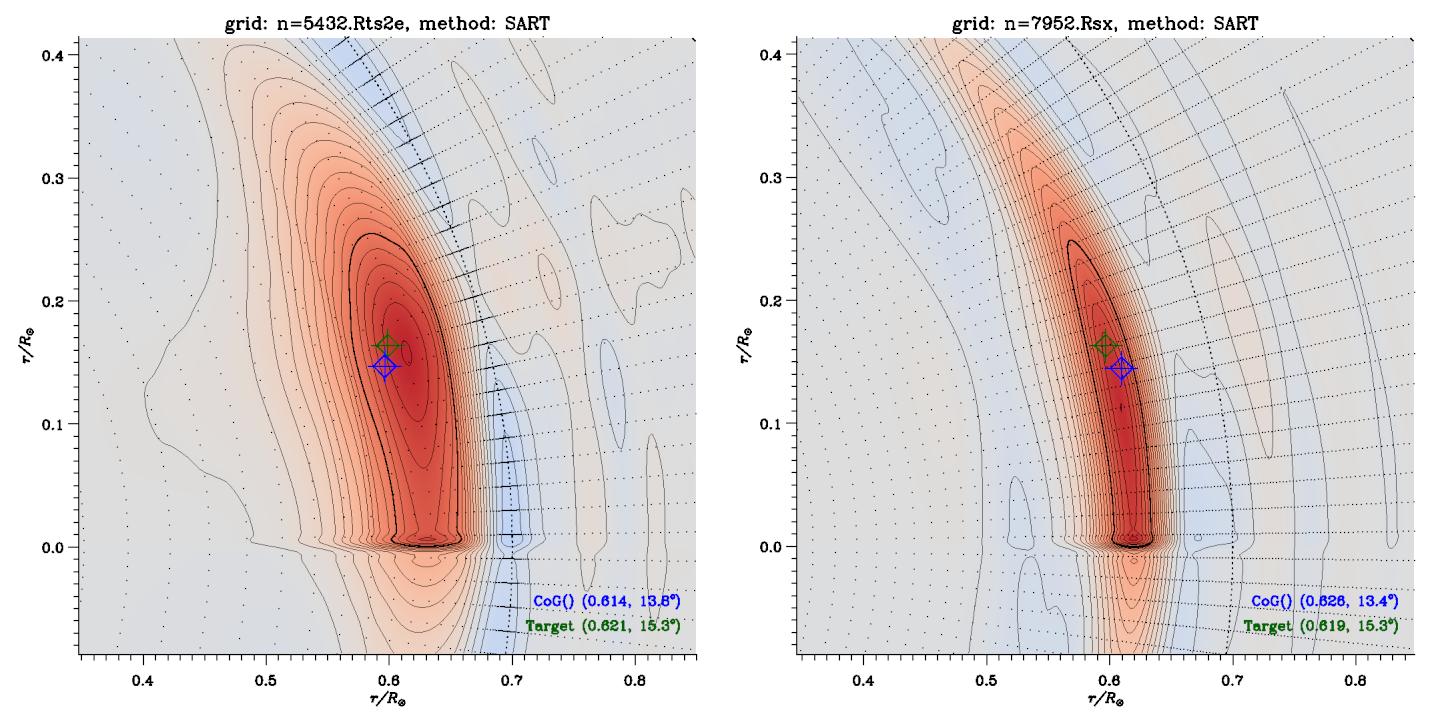}}} \\ 
    {\tiny ~~~~(e) $\phi=15^{\rm o}$ RLS \hspace*{.30\textwidth} (f) SART \hfill ~}
  \end{center}
  \caption{Animations of average kernels at three latitudes ($\phi=0, 7.5\, \&
    15$\textdegree), for two grids and two inversion methodologies.  The
    animations show the evolution of the average kernels as the target radius
    changes, comparing the {\em aggressive} with the {\em progressive I} grid
    (left versus right panels).  Notice how the average kernel is ``jumping''
    as the target radius crosses the tachocline when using the {\em
      aggressive} grid, but not the {\em progressive} one.}
  \label{anim:kernels}
\end{figure}

This being said, we recognize that using the center of gravity, even from high
resolution extended averaging kernels is most likely not the optimal way to
determine the effective location of the derived solution. Yet it also shows
how when using the aggressive grid this effective location is skewed by the
averaging kernels.  This is best illustrated by the animations in
Fig.~\ref{anim:kernels}, that show how, at some fixed latitude, the averaging
kernels change as a function of radius around the tachocline, for different
inversion grids and inversion methodologies.

These animations show how for the {\em aggressive} grid, the location of the
averaging kernel, and hence its center of gravity, jumps around for the target
locations near the tachocline where the radial density is very high. In fact
the center of gravity, indicated by the blue symbol, loops around the target
location. This effect is more pronounced when using the RLS method, but
disappears when using the {\em progressive} or {\em equispaced} grids for
either inversion technique.

\begin{figure}
  \begin{center}
    \includegraphics[width=.85\textwidth]{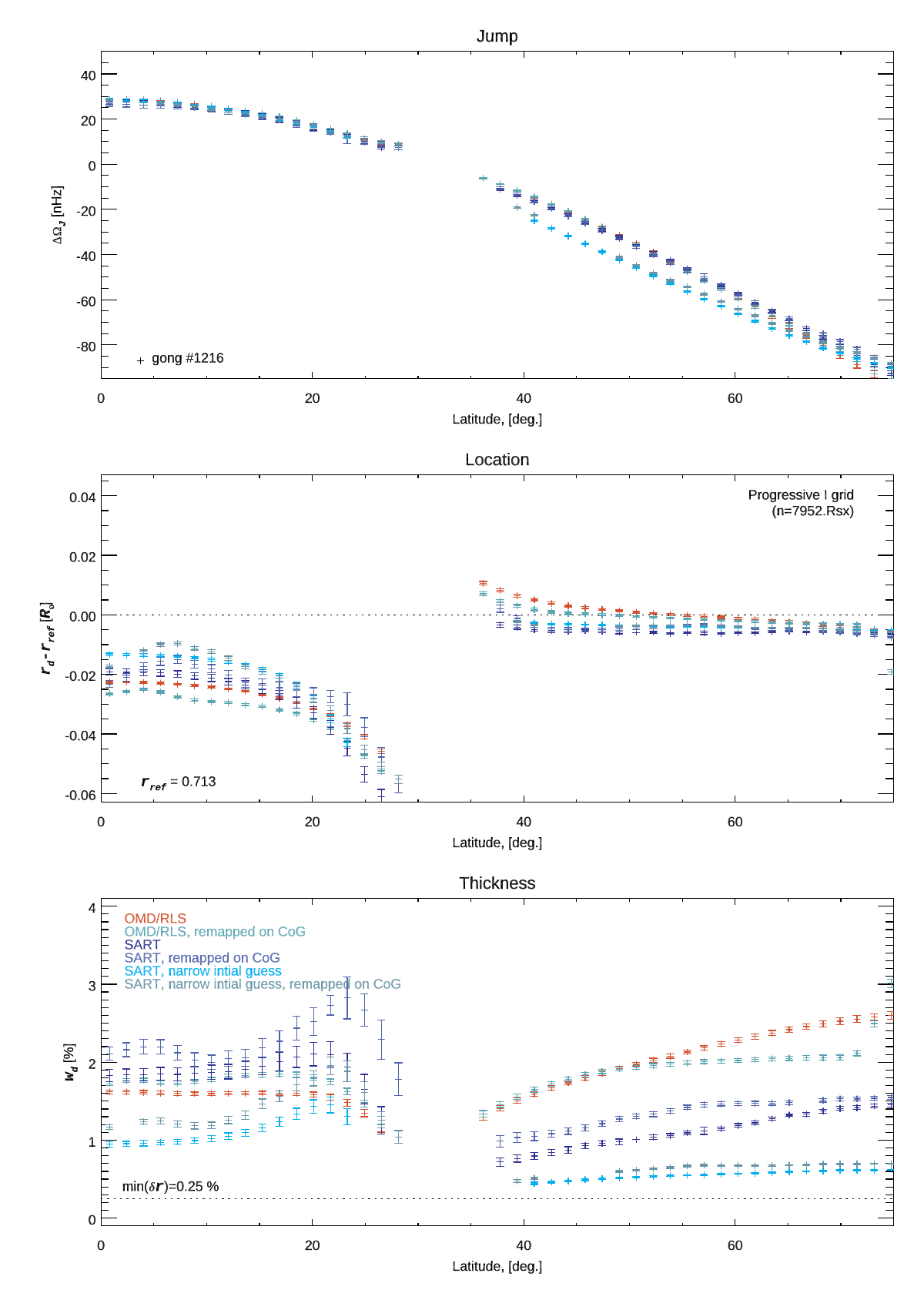} 
  \end{center}
  \caption{Characteristics of the tachocline resulting from fitting a sigmoid
    to inverted rotation rates derived from fitting a very long time series,
    i.e., 25.2 year of GONG observations, plotted as a function of latitude,
    after remapping the rotation rate on the center of gravity of extended
    high resolution average kernels, or not. It shows that the remapping
    affects mostly our estimates of the tachocline thickness, the parameter
    most difficult to constrain, but does not significantly change our
    estimates of its position.}
  \label{fig:sigmoid-results-128e-on-cog}
\end{figure}

With this caveat in mind, we remapped the rotation rate on the center of
gravity of the extended high resolution averaging kernels and fitted our
sigmoid model.  The resulting parameters are shown in
Fig.~\ref{fig:sigmoid-results-128e-on-cog}, and yet despite some small changes
in the resulting estimates, the main characteristic of these estimates remains
unchanged, including the discontinuity of the tachocline location, $r_d$.

\begin{figure}
  \begin{center}
    \includegraphics[width=.8\textwidth]{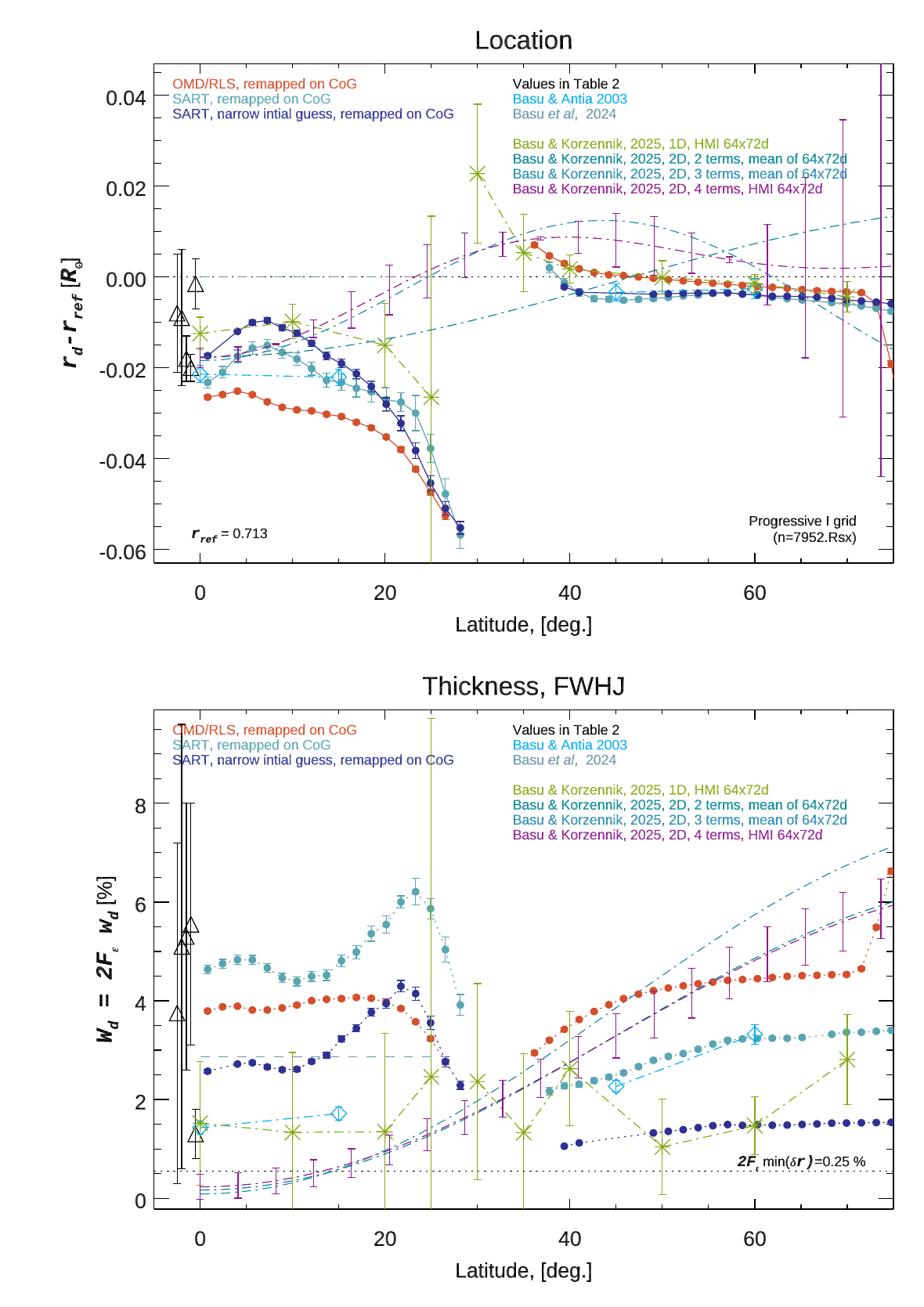} 
  \end{center}
  \caption{Characteristics of the tachocline, \ie, position and FWHJ, as a
    function of latitude resulting from fitting a sigmoid to inverted rotation
    rates derived from fitting a very long time series, \ie, 25.2 year of GONG
    observations (filled circles), for three cases (see legend).  These are
    compared to the equatorial values in Table~\ref{tab:rot-tacho-sum} (black
    triangles, offset to negative latitudes for clarity) and to values
    estimated using a forward modeling technique, from \citet{Basu+Antia-2003}
    (diamonds), from the more recent \citet{BasuEtal-2024} (horizontal dash
    line). It also show results from \citep{Basu+Korzennik-2025}, whether
    using a one dimensional or a two-dimensional model (star vs.\ dot-dash
    lines), and for the latter weather using 2, 3 or 4 terms in the polynomial
    expansion of $r_t$ \citep[see legend and ][for additional
      details]{Basu+Korzennik-2025}. }
  \label{fig:sigmoid-results-128e-comp}
\end{figure}

Next, we plot on Fig.~\ref{fig:sigmoid-results-128e-comp}, the tachocline
location and width (\ie, the FWHJ) derived from three inversions (RLS, SART
and SART with narrow initial guess, all remapped on the effective locations,
using one of the progressive radial density grids) and compare them to other
estimates.  Namely to the equatorial values tabulated in
Table~\ref{tab:rot-tacho-sum}, but also to the values reported by
\citet{Basu+Antia-2003} and more recently by \citet{BasuEtal-2024}, values
computed using forward modeling of the splitting expansion
coefficients. \citet{Basu+Antia-2003} used one and two-dimensional rotation
models and the respective projects pipeline results (GONG and MDI) derived
from fitting 108 or 72 day long time series. The plotted values are the
average values computed from the non-overlapping GONG and MDI data set
spanning the 1995 to 2002 interval.
The more recent \citet{BasuEtal-2024} results are based on splitting estimated
from much longer time series \citep{Korzennik-2023} but uses a one dimensional
rotation model, hence produced equatorial estimates, although effectively
these estimates correspond to some average over a range of latitudes around
the equator (indicated by the latitudinal extent of the corresponding
horizontal lines).

This forward modeling, derived from the same splittings, has since been
expanded to either a one dimensional model at various latitudes or a
two-dimensional model of the tachocline using a polynomial expansion with 2, 3
or 4 terms for the latitudinal variation of $r_d$
\citep{Basu+Korzennik-2025}. Results from that work are also indicated on
Fig.~\ref{fig:sigmoid-results-128e-comp}.  Note that the formal uncertainties
on the tachocline characteristics derived from fitting a sigmoid model to
inversion inferences are bound to be small, since the inversion profile itself
is a smoothed estimate of the true underlying rotation profile.

Note the remarkable agreement between most methods, although using a
polynomial expansion for the forward modeling dependence on latitude cannot
reproduce the discontinuity seen in our two-dimensional inversions. Attempts
to carry out forward modeling with more terms or by including a discontinuity
has so far produced worse models with excessively large uncertainties at high
latitudes \citep[see][]{Basu+Korzennik-2025}.

\begin{figure}
  \begin{center}
    \includegraphics[width=.425\textwidth]{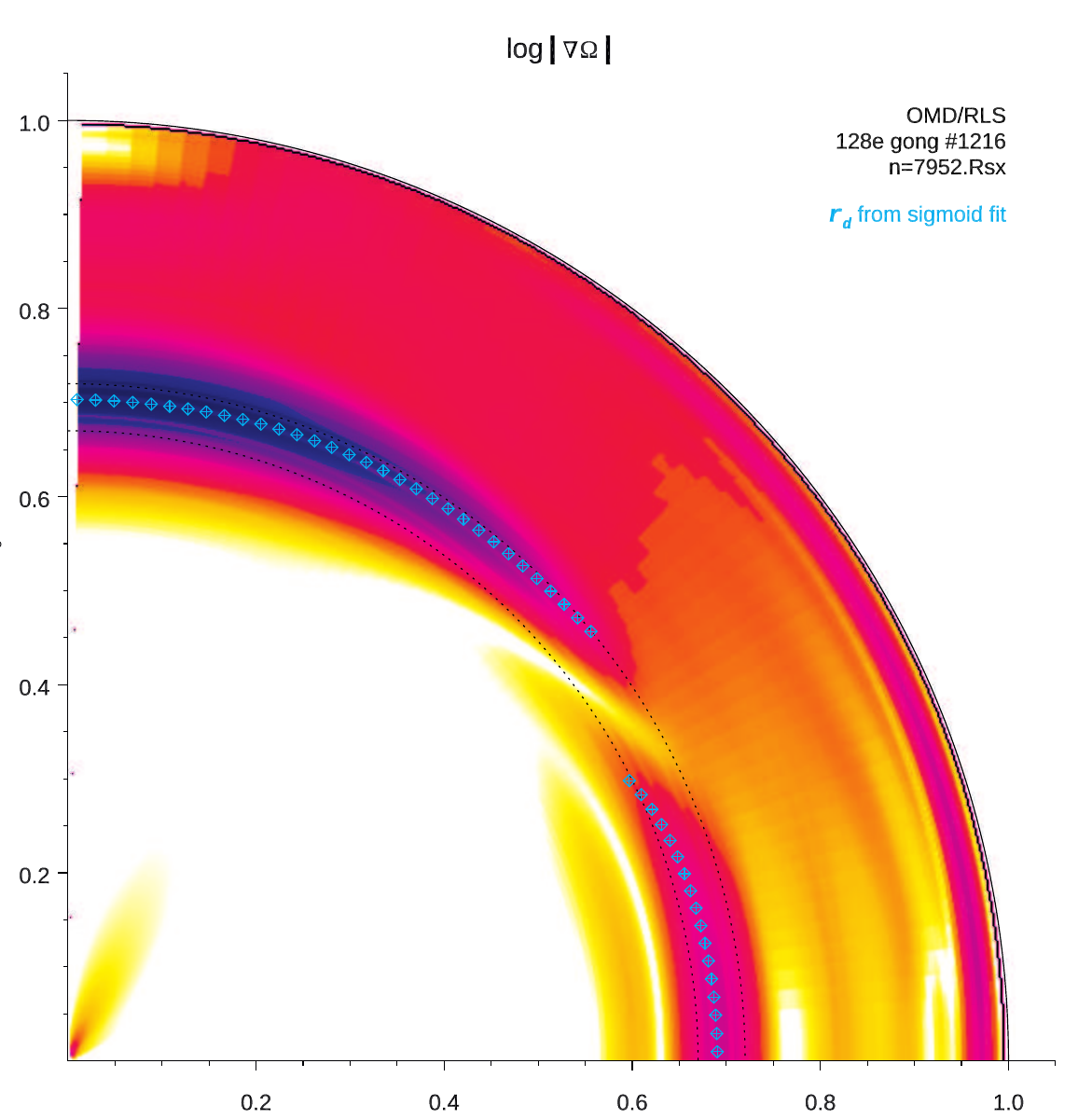} 
    \includegraphics[width=.425\textwidth]{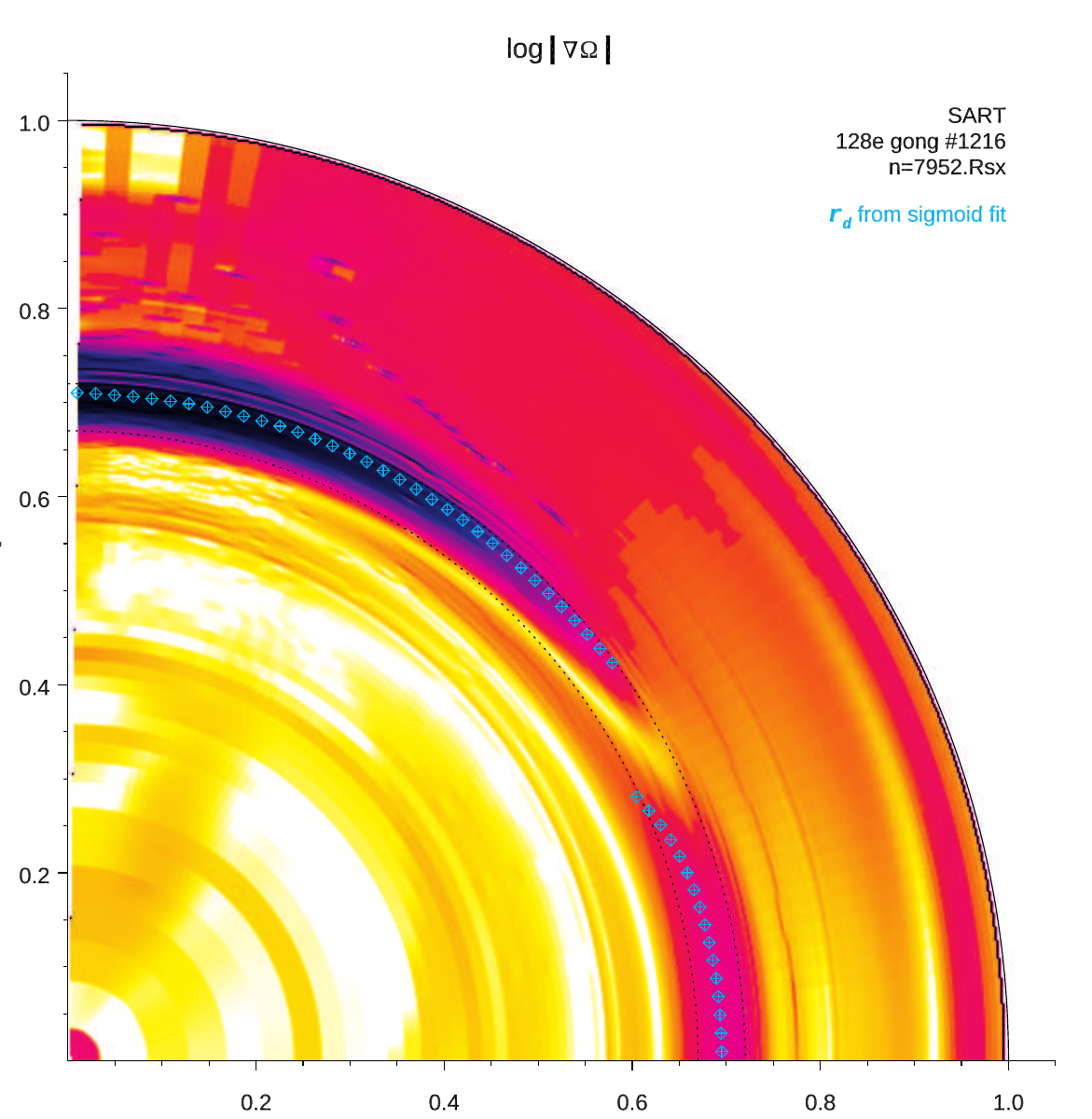} 
  \end{center}
  \caption{Amplitude of the rotation rate gradient, $|\nabla\Omega|$,
    resulting from inverting splittings derived from a very long time series
    and using either inversion methodology (left versus right panel). The
    gradient is shown on a logarithmic scale and the position of the jump
    location, $r_d$, measured by fitting a sigmoid is shows as well
    (crosses).}
  \label{fig:nabla-omega}
\end{figure}

Another way to look at the location of the tachocline is to compute the
rotation rate gradient, and visualize its amplitude, namely
$|\nabla\Omega|$. Fig.~\ref{fig:nabla-omega} shows the amplitude of this
gradient and compares it to the location of the tachocline derived from
estimates of $r_d$.  These estimates match the location of the maximum
gradient and while the gradient derived using the SART methodology is more
noisy, the visualization of the gradient shows clearly the duality of the
location of the tachocline, curving inwards with latitude at low latitudes
only to reappear at larger radii at the higher latitudes.

\subsection{Tachocline Characteristics Using Long Time Series}

We also inverted the splittings resulting from fitting somewhat shorter time
series, namely 12.6 and 6.3 year long observations of GONG, MDI and HMI. Our
sigmoid model was then fitted to the inferred rotation rates. Since fitting
shorter time series leads to larger uncertainties, the smoothing was somewhat
increased in each case compared to the smoothing used for the very long time
series.

\begin{figure}
  \begin{center}
    \includegraphics[width=.975\textwidth]{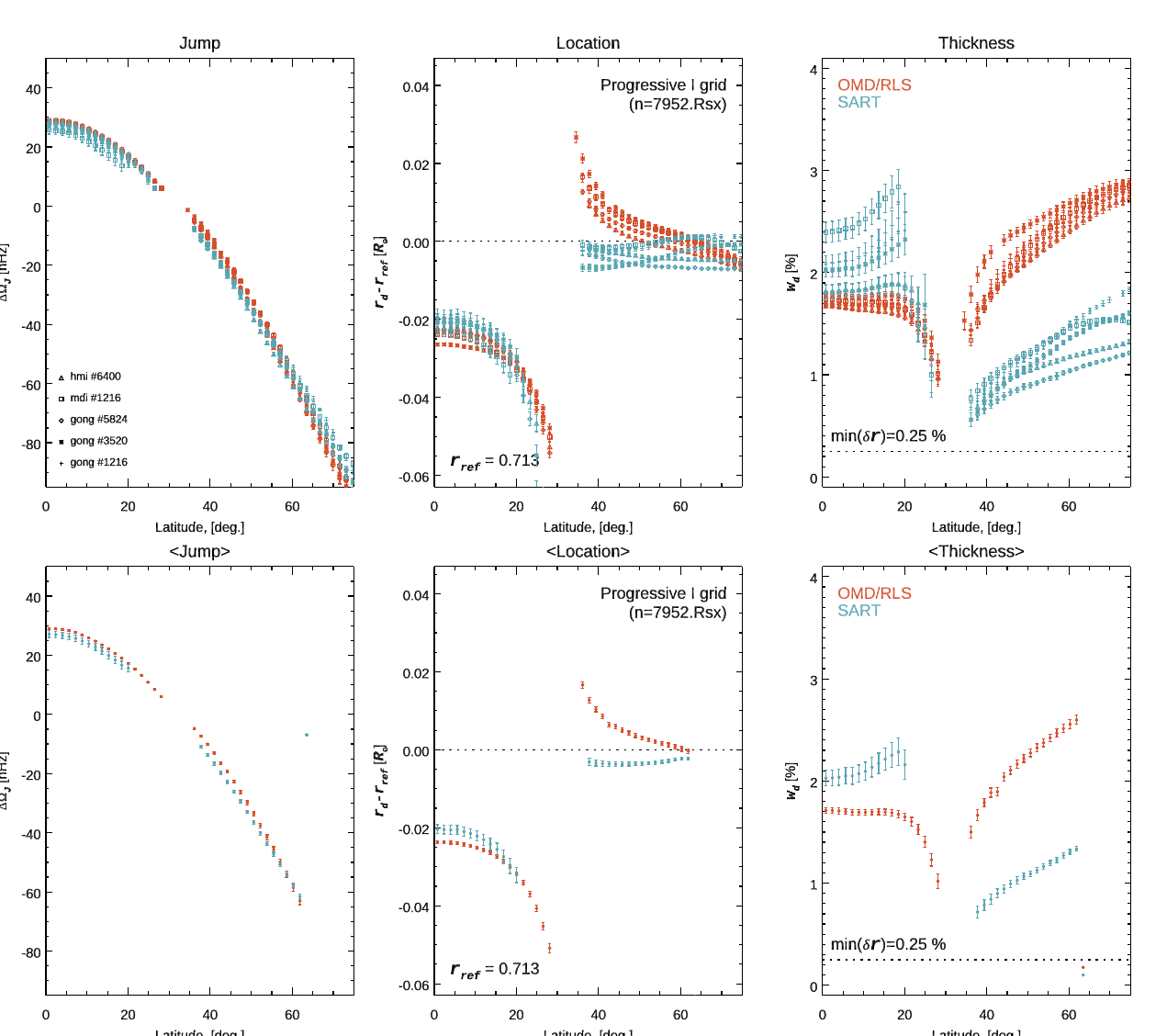} 
  \end{center}
  \caption{Top panels: characteristics of the tachocline resulting from
    fitting a sigmoid to inverted rotation rates derived from fitting a 12.6
    year-long time series of GONG, MDI and HMI observations, i.e., six sets of
    splittings and both methodologies.  The jump, location and width are
    plotted as a function of latitude, with colors indicating the inversion
    method, while the different symbols correspond to the different splittings
    sets.  The bottom panels show, for each method, the mean jump, location
    and width plotted as a function of latitude, with colors indicating the
    inversion method.  }
    
  \label{fig:sigmoid-results-64e}
\end{figure}

\begin{figure}
  \begin{center}
    \includegraphics[width=.975\textwidth]{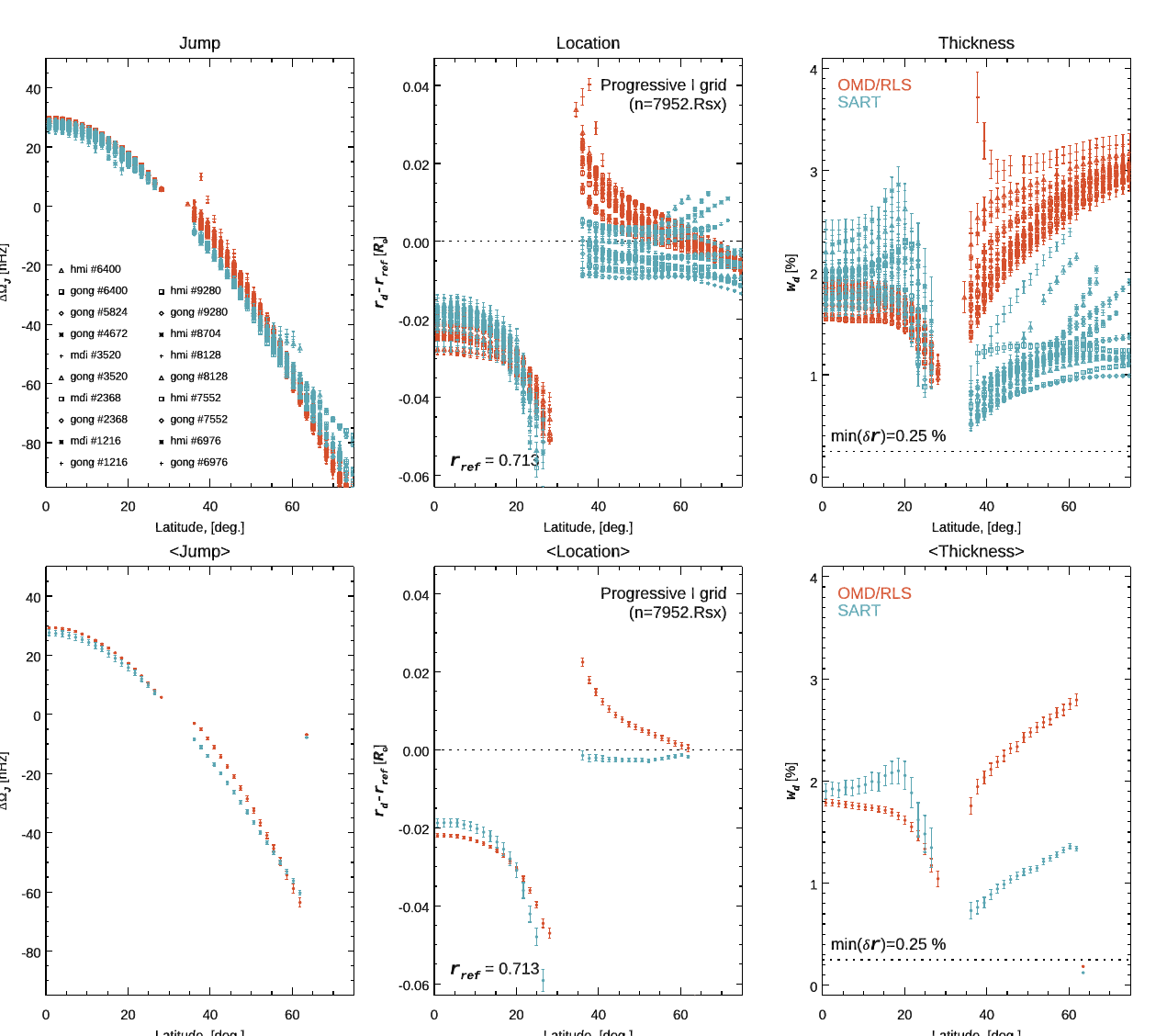} 
  \end{center}
  \caption{Top panels: characteristics of the tachocline resulting from
    fitting a sigmoid to inverted rotation rates derived from fitting a 6.3
    year-long time series of GONG, MDI and HMI observations, \ie, 19 sets of
    splittings and both methodologies.  The jump, location and width are
    plotted as a function of latitude, with colors indicating the inversion
    method, while the different symbols correspond to the different splittings
    sets.  The bottom panels show, for each method, the mean jump, location
    and width plotted as a function of latitude, with colors indicating the
    inversion method.  }
  \label{fig:sigmoid-results-32e}
\end{figure}

Figs.~\ref{fig:sigmoid-results-64e} and \ref{fig:sigmoid-results-32e} show the
resulting tachocline characteristics, when using either inversion method on
each available data set. While each method returns slightly different values,
especially at higher latitudes, the discontinuity of the location of the
tachocline is consistently observed for each data set. Not surprisingly the
RLS method returns a wider tachocline at high latitudes, something that we
attribute to the method greater smoothing. The mean characteristics, derived
from averaging over the data sets, are also plotted in these figures and are
consistent with each other and the corresponding characteristics derived from
the very long time series.

\begin{figure}
  \begin{center}
    \includegraphics[width=.975\textwidth]{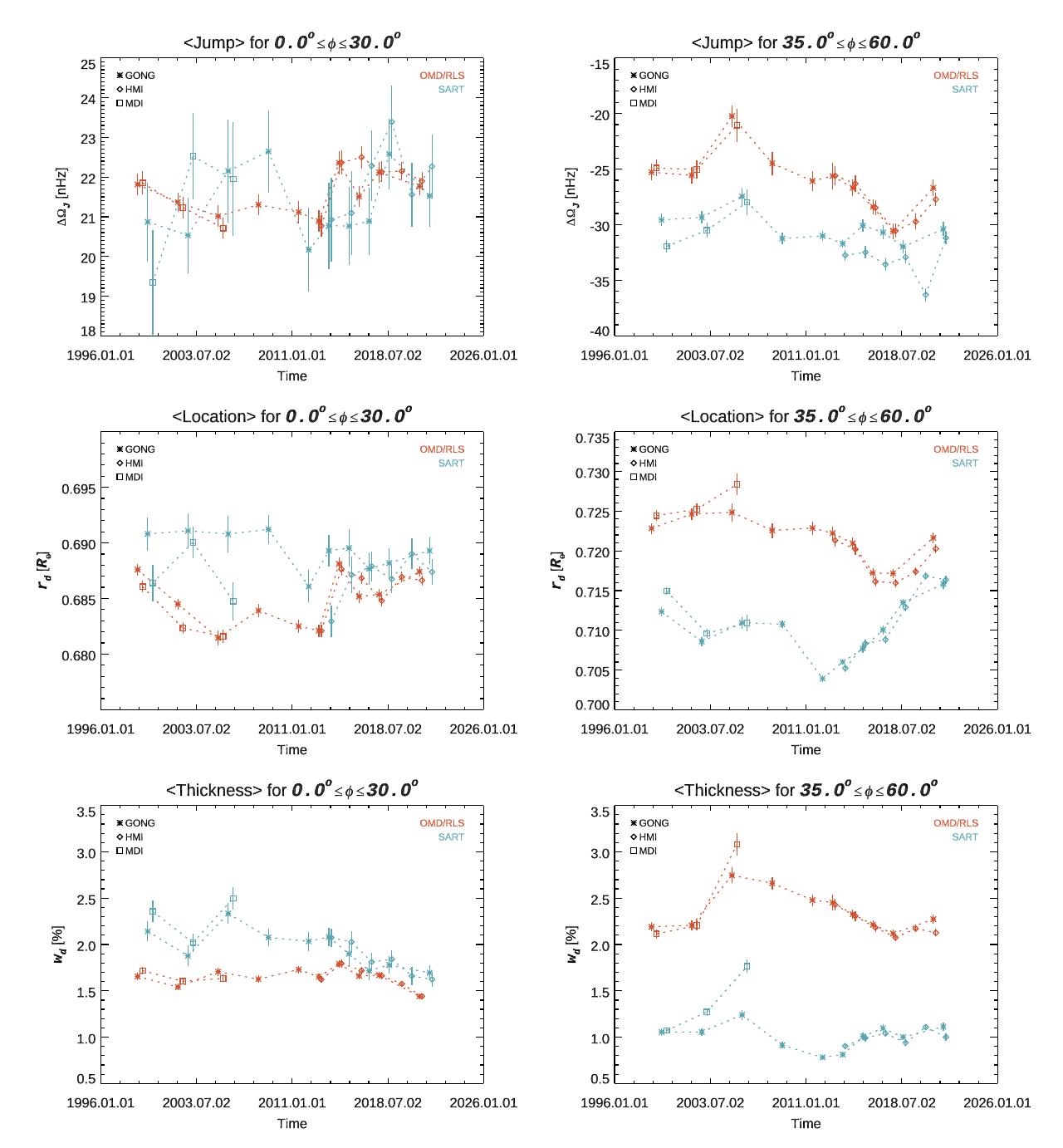} 
  \end{center}
  \caption{Characteristics of the tachocline resulting from fitting a sigmoid
    to inverted rotation rates derived from fitting a 6.3 year-long time
    series of GONG, MDI and HMI observations, \ie, 19 sets of splittings,
    using either inversion methodology.  These were averaged over two ranges
    in latitude and plotted as a function of time.  The colors correspond to
    the inversion method, while the different symbols correspond to the
    different data sets (GONG, MDI, HMI).  The panels on the left show the
    mean values for low latitudes ($\phi\le30^{\rm o}$), while the panels on
    the right show the mean values for high latitudes ($40^{\rm
      o}\le\phi\le70^{\rm o}$).  These results are too noisy to detect a
    definite or significant temporal variation, although some trends with time
    are present.  }
  \label{fig:sigmoid-vs-time}
\end{figure}

Temporal changes of either the tachocline jump amplitude, location, or width
derived from fitting a sigmoid model to inversion results, even when averaged
over either low or high latitudes (\ie, for $\phi \le 30^{\rm o}$ or for
$35^{\circ} \le \phi\le 60^{\circ}$), remain quite noisy, as shown in
Fig.~\ref{fig:sigmoid-vs-time}. Despite the latitudinal averaging there is no
definite or significant temporal variation, although the scatter and the
formal error bars are somewhat smaller for inferences using the RLS inversion
methodology.
At low latitudes these means indicate a possible decrease of the jump
amplitude by about 1 nHz around 2006, a trend that is not quite compatible
with the changes inferred in \citet{BasuEtal-2024}, where such decrease peaks
around 2002. These also suggest a correlated small change in the location,
\ie, a tachocline located a little deeper, by $\sim 0.005 R_{\odot}$, when the
amplitude of the jump is smaller.
The figure also shows a marginal decrease of the mean width with time at low
latitudes, but the same characteristics derived when using the SART method
(jump, position and width), although more scattered and noisy, do not align
with the RLS method results.
At high latitudes, the inferred jump amplitude appears to also change with
time, but here again the changes are method dependent although more
pronounced.
These systematic differences are most likely the result of the different
amount of effective smoothing of each method, especially at high latitude.

\section{Conclusions} \label{S-Conclusions}

Using very long and long time series, we have carried out rotation inversions
using two different methodologies. Since we wanted to characterize the
tachocline we have first shown how each method performs when adjusting the
inversion grid, especially in the radial direction around the tachocline and
when fine tuning each method's smoothing and trade-off. To do this we used
simulations without and with random noise. We conclude from this that one
should not use an aggressive radial grid, \ie, a grid that becomes very dense
{\em only} around the tachocline, as it produces artifacts, especially once
noise is included.
Yet, using a radial grid with a progressive transition of its radial density
mitigates most of these artifacts, although it requires more computing
resources, especially if one aims to reach a very high radial resolution at
the tachocline.
We also have shown that one can improve the inferences of the SART method by
using an initial guess derived from the RLS solution, although fine tuning
both the way that initial guess is constructed (\ie, sharpened) and the
precise smoothing parametrization of the SART method remains delicate.

By fitting a sigmoid model of the tachocline to our inverted profiles we
inferred the variation with latitude of the rotation radial shear, \ie, the
amplitude of the rotation rate jump across the tachocline, the location of
this shear and an estimate of the width of the tachocline.

We find that the location of the tachocline does not vary smoothly with
latitude, but instead shows a clear discontinuity between its position at low
latitudes, where the radial shear is positive, and its position at high
latitudes, where that shear is negative. This is also clearly seen when
looking at the rotation rate gradient and persists when using splittings
derived from somewhat shorter time series and when using either inversion
methodology.

Also, our estimates of the position and its variation with latitude match
rather well estimates based on forward modeling using rotational splitting
coefficients as in
{\citet{Basu+Antia-2003, BasuEtal-2024} and figures 5 \& 6 of the more recent
  \citet{Basu+Korzennik-2025}}.
Of course the discontinuity cannot be seen when such modeling uses a
polynomial expansion of the tachocline characteristics with respect to
latitude.
We find that the tachocline lies below the convection zone, by
$0.02R_{\odot}$, at low latitudes ($\phi \le 30^{\circ}$) only to coincide
with the convection zone at the higher latitudes, once the radial shear is
negative.

{While we have made every effort to eliminate the possibility that the
  bifurcation we observe in the tachocline could be an inversion artifact,
  \ie, the result from limited latitudinal resolution, kernel geometry, or
  fitting degeneracies, especially when the jump amplitude becomes small, this
  is a new paradigm for the shape of the tachocline and thus would benefit
  from independent confirmation. Not suprisingly, we found no modeling of the
  tachocline that would produce or explain such bifurcation.}

Our estimate of the tachocline width is commensurable with previous estimates
and suggests that its FWHJ might be as low as 1\% of the solar radius, if not
lower {since rotation inversion inferences are limited by the resolution
  of the averaring kernels}.
{Estimates of this width when using a narrow initial guess for the SART
  case could be biased were the data a lot more accurate, hence allowing for
  much less smoothing. The values derived are overall in good agreement with
  the values derived by others, including \citet{Basu+Korzennik-2025}, as
  shown in Fig.~\ref{fig:sigmoid-results-128e-comp}, despite the scatter.}
Yet, our inferences of its variation with latitude are still very method
dependent, hence we cannot ascertain whether it really increases with
latitude, as seen in inferences from forward modeling, {although recent
  work using forward modeling of rotational splittings coefficients, using
  long time series of GONG observations (720-day long) and two different mode
  fitting pipe-lines show similar trends \cite{BasuEtal-submitted}}
Also, our estimate of the variation of the amplitude of the radial shear with
latitude agrees rather well with corresponding estimates derived from forward
modeling.

We introduced the concept of the extended averaging kernel, since we
checked that our inferences of the tachocline characteristics were not an
artifact of the variation of the effective position of the inferred
rotation rate. Indeed, inversion inferences are a smoothed representation 
of the underlying {\em true} rotation rate, although we must add that the 
complex nature of the averaging kernel makes it tricky to derive an effective 
position.

As for changes with time, hence solar activity, we find that our current
rotation inversion implementations remain too noisy to see definitive and
significant temporal variations. We recognize that there remain systematic
differences when using different methodologies.  {These method-dependent
  systematics currently dominate the resulting temporal variations, especially
  at high latitudes.}  We plan to further improve this by devising a more
objective selection of the amount of smoothing and inversion trade off between
resolution and error magnification for both methods.

\clearpage

\begin{acknowledgements}

  This work was partially supported by NASA grants 80\-NSSC22K0516 and
NNH18ZDA001N--DRIVE to SGK and by the Spanish AEI programs
ID2022-–139159NB–-I00 (Volca-Motion) and PID2022–-140483NB-–C21 (HARMONI) to
AED.

\end{acknowledgements}
\end{document}